# PolyGPT: A Decoder-Only Transformer for Polymer Electrolyte Design

Tanmay Sarkar Akash, Siddhartha Das*
*University of Maryland*
*Corresponding author

## Abstract

Designing polymer electrolytes requires control over both material properties and chemical structure. Here, we present PolyGPT, a Transformer with 38.3 million parameters that generates polymer repeat units from optional property targets, requests to include or exclude a specified chemical feature, and an electrolyte-domain condition. Using the model selected by validation loss, we compared unconditional generation with requests for electrolyte-associated structures, a low glass-transition temperature (Tg), or both. Unconditional and electrolyte-only generation achieved structural validity of 99.46% and 99.59%, respectively. However, electrolyte conditioning increased repetition: uniqueness among valid outputs fell from 99.99% to 51.74%. At a Tg target of 246.11 K, we compared generation with and without electrolyte conditioning, and adding the electrolyte condition slightly reduced the predictor-estimated mean absolute error from 26.97 to 26.07 K, but the fraction of all attempts yielding distinct valid polymers within 50 K of the target fell from 82.35% to 32.31%. When we also required absence from the training catalog, these yields were 59.90% and 19.71%, respectively. The lower yield persisted with a tighter 25 K tolerance. Domain conditioning also weakened control of inclusion requests: the mean fraction of attempts producing valid polymers containing the requested chemical feature fell from 94.27% to 31.27%. Across nine checkpoints, earlier models sometimes produced more distinct, catalog-unseen polymers meeting the Tg target: 10,000 checkpoint model with electrolyte conditioning and 30,000 checkpoint model without it. Together, these results show why property error alone is insufficient for evaluating conditional polymer generation. It must be considered alongside repetition, training-catalog novelty, and whether combined design requests are satisfied.

## Introduction

Polymer electrolytes offer a broad chemical space for tuning ion transport, thermal response, and interactions with dissolved salts. Exploring this space is difficult because a useful electrolyte must satisfy several requirements at once. A repeat unit that promotes segmental motion may differ from one that provides favorable ion coordination, and the conductivity of the resulting material also depends on its formulation and operating temperature. Data-driven polymer design can help identify candidates for this coupled problem by connecting structural proposals with estimates of relevant properties [1, 2, 3]. The central challenge is to generate structures that meet a specified combination of requirements while retaining enough chemical diversity to support discovery.

Machine learning has expanded both property prediction and inverse design in polymer informatics. TransPolymer, polyBERT, and PolyNC learn relationships between polymer representations and material properties, providing tools for evaluating large candidate collections [4, 5, 6]. Generative approaches address a complementary task by proposing structures for a desired region of property space. Syntax-directed variational autoencoders, invertible molecular graph models, and graph-based methods such as polyG2G have demonstrated routes to polymer design beyond exhaustive enumeration of a fixed library [7, 8, 9]. Their performance depends on the representation of the polymer, the information supplied to the model, and the criteria used to select its outputs.

Chemical language models can condition polymer-sequence generation on specified design requirements. PolyTAO learns property-to-structure mappings for on-demand generation, while polyBART and POLYT5 use encoder-decoder models for polymer prediction and design [10, 11, 12]. Decoder-based approaches offer another formulation. PolymerGPT incorporates multiple numerical properties through learned conditioning prefixes and also supports scaffold constraints [13]. In molecular generation, Llamol demonstrates how optional numerical and structural contexts can be accommodated within one autoregressive model [14]. These approaches motivate testing how numerical and structural conditions interact during generation.

Polymer electrolyte design provides a useful setting for this question because application-specific data cover a much smaller and chemically different region than broad hypothetical-polymer collections. GPT-based and diffusion-based electrolyte generators have been evaluated through conditional sampling and atomistic simulations, and conditional generative models have also been incorporated into iterative discovery frameworks [2, 3]. These studies connect candidate generation with independent evaluation of transport. In a more general polymer model, an electrolyte-domain condition can instead identify the distribution associated with an application dataset. Such a condition may favor relevant chemistry, but its effect on a numerical target must be determined from the generated structures.

Several aspects of joint control are especially important for this application. Numerical conditions describe desired properties, whereas structural conditions specify chemical features that should be present or absent. An electrolyte-domain condition instead indicates that a polymer belongs to the electrolyte-associated training collection. These requests can either reinforce each other when their combination is well represented in the training data or compete with each other when it falls outside the available support. Structural validity alone does not show whether the generated candidates satisfy these combined requirements. For screening, a more useful measure is the fraction of all generation attempts that produce distinct polymers satisfying every requested condition. Training-catalog novelty provides a further, separate measure of whether those candidates extend beyond known source representations [15, 16].

Here, we develop PolyGPT to study these interacting controls within a single conditional autoregressive decoder-only Transformer. The model accepts nine optional numerical channels, one inclusion or exclusion request chosen from eleven chemical annotations, and an optional electrolyte-domain token. We first establish generation quality and chemical diversity, then evaluate numerical and structural control separately and in combination. A comparison of nine model checkpoints at equal sample sizes examines whether continued training improves the yield of distinct candidates that are absent from the training catalog and satisfy the requested conditions.

## Results and discussion

### PolyGPT framework

PolyGPT represents polymer design as conditional sequence generation (Figure 1). The model combines requested numerical properties and indicators of which properties were specified into a conditioning vector. Separate optional tokens identify electrolyte association and a structural request. The indicators identify which property targets are specified, allowing generation with no numerical targets, one target, or several targets. The decoder uses this information to generate a polymer repeat sequence one token at a time. Structural requests specify the inclusion or exclusion of one annotated feature; they do not require an exact scaffold to be reproduced.

We trained the model jointly on a broad polymer corpus and a smaller electrolyte dataset. The prepared foundation dataset contains approximately 102.5 million eligible records, while the electrolyte dataset contains 5,935 groups formed using the conservative polymer-identity procedure. To limit overrepresentation of the smaller electrolyte collection, we restricted how often each electrolyte polymer group could be sampled during training. Training stopped at 49,000 optimizer updates when validation loss no longer improved. The model at 43,000 updates had the lowest balanced validation objective and was selected before comparing generated structures across checkpoints; it supplies the principal results (Figure S6). By this stage, the model had processed 44.032 million training examples, including repeated sampling, and 3.669 billion supervised sequence tokens. Architecture, data partition, and training details are given in Methods and Supporting Information Section S1.

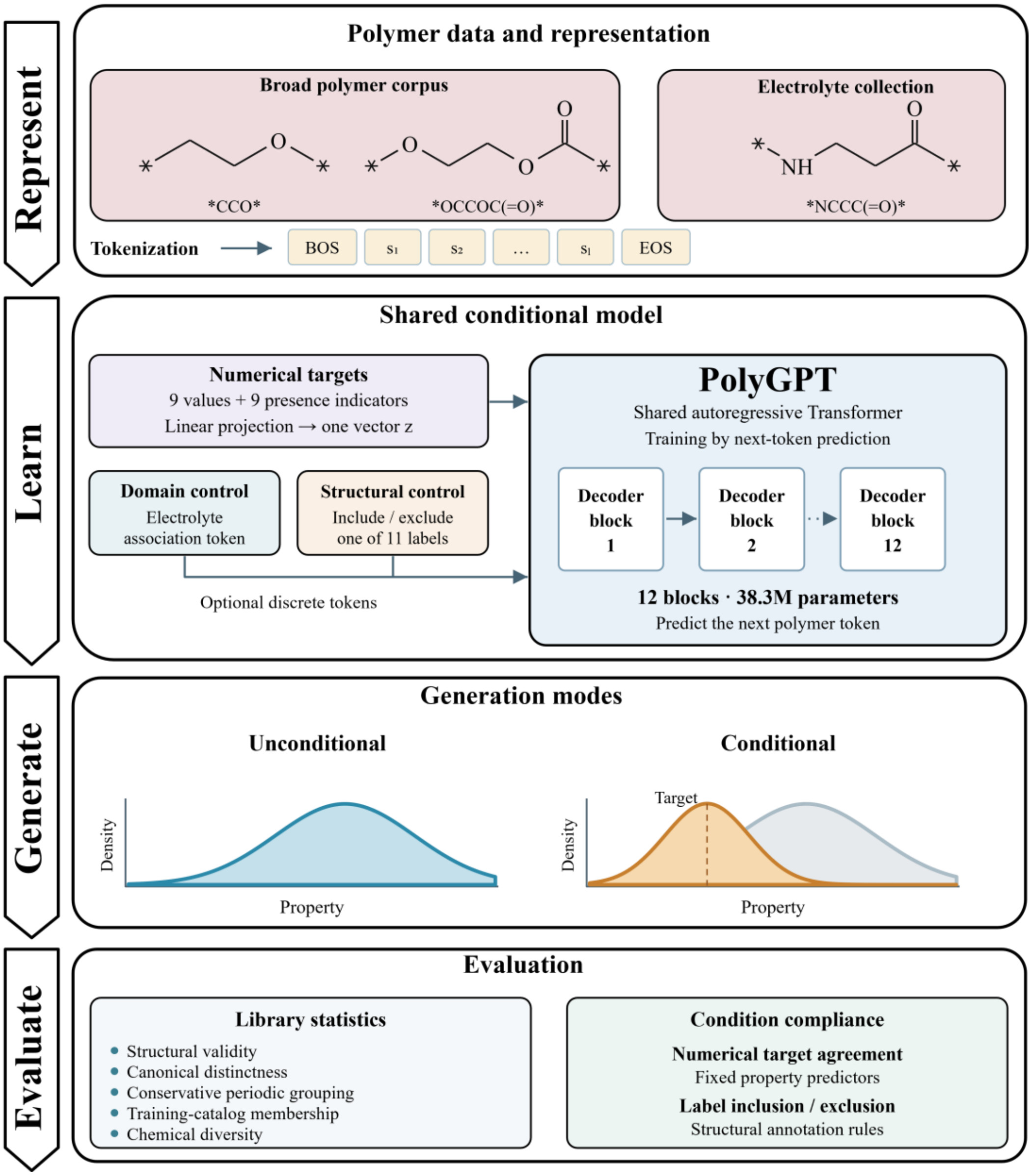

**Figure 1.** PolyGPT framework. **Represent**: repeat-unit sequences from the foundation and electrolyte-associated collections are tokenized; the depicted structures are illustrative. BOS and EOS denote the beginning-of-sequence and end-of-sequence tokens, respectively, and $s_i$ denotes a polymer-sequence token. **Learn**: nine optional numerical values and presence indicators are projected into one conditioning vector. Optional discrete tokens specify electrolyte association and inclusion or exclusion of one of eleven structural annotations. A shared 12-block decoder generates repeat units. **Generate**: schematic distributions illustrate unconditional and conditional generation and are not quantitative results. **Evaluate**: assessment covers library statistics, predictor-based numerical agreement, and rule-based label compliance.

After selecting the model, we tested it on 5,074 held-out examples containing 428,226 tokens. The average token-prediction loss was 0.20728. The balanced test loss was 0.24621, calculated by equally weighting the validation categories described in Supporting Information Section S1.3. These scores measure how well the model predicts polymer sequences, not the physical properties of generated polymers (Table S20 and Supporting Information Section S5).

Fixed property predictors assess agreement with numerical requests, while structural annotations assess inclusion and exclusion. The electrolyte token indicates that the source polymer was associated with an electrolyte dataset; it does not request a particular conductivity. We therefore evaluate its effect on chemical distributions and target compliance without assigning transport performance to the generated structures.

**Generation quality and chemical diversity**

We first evaluated the selected model under four conditions: unconditional, electrolyte-only, low-Tg-only, and low Tg with electrolyte control (Figure 2 and Table 1). All invalid and repeated outputs were retained in the attempt denominator. Unconditional generation achieves 99.46% validity and a distinct canonical yield of 99.45%. With electrolyte conditioning, validity is 99.59%, but distinct canonical yield falls to 51.53%. Validity therefore remains high while duplication becomes substantial at the library scale.

**Table 1.** Pooled generation quality for the selected 43,000-update model. Uniqueness is the number of distinct canonical repeats divided by valid generations; novel distinct yield is the number of distinct canonical repeats absent from the training catalog divided by all attempts. Periodic counts use the conservative identity procedure, which does not establish complete repeat equivalence. Low Tg denotes 246.11 K.

| Condition | Valid | Canonical distinct | Periodic distinct | Unique (%) | Novel distinct | Novel yield (%) |
|---|---|---|---|---|---|---|
| Unconditional | 29,837 | 29,835 | 29,832 | 99.99 | 13,189 | 43.96 |
| Electrolyte | 29,876 | 15,458 | 12,881 | 51.74 | 8,745 | 29.15 |
| Low Tg | 29,774 | 28,588 | 28,326 | 96.02 | 20,621 | 68.74 |
| Low Tg + domain | 29,849 | 11,664 | 9,580 | 39.08 | 6,964 | 23.21 |

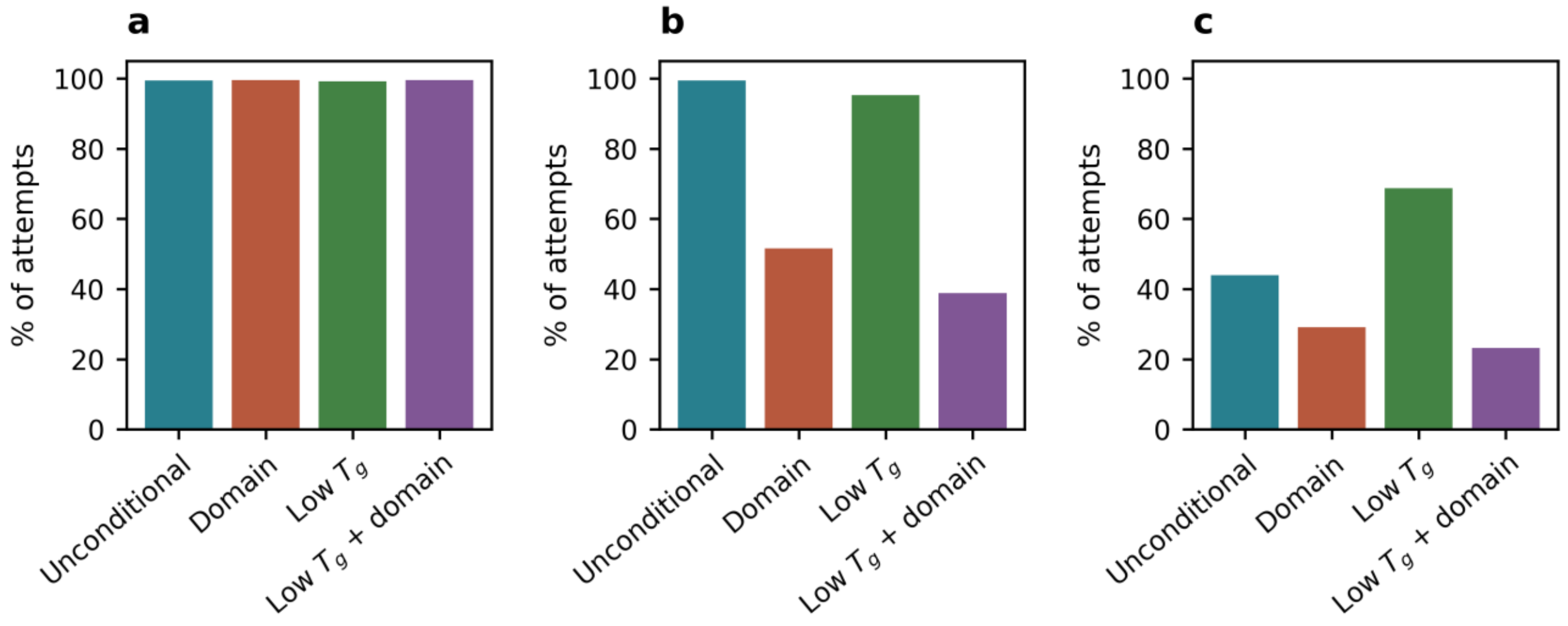

**Figure 2.** Pooled generation quality in the principal evaluation: (a) structural validity, (b) canonical distinct valid yield, and (c) yield of distinct structures absent from the training catalog. Distinct and novel distinct counts are computed after pooling and deduplicating generations from the three random seeds. Individual-seed results and their variation are reported separately in Table S17.

The novel distinct yield is 43.96% for unconditional generation and 29.15% with electrolyte conditioning. Low-Tg-only generation yields 68.74%, compared with 23.21% under the combined low-Tg and electrolyte request. These yields measure absence from the complete eligible training catalog, not novelty relative to all known experimental chemistry. Catalog membership also does not prove memorization, because the model may have seen other records during training.

Figure 3 examines how the four generation conditions change the structural distributions and their similarity to the structures actually encountered during training. Complementary size and internal-diversity results, including the three individual sampling seeds, are reported in Tables S21 and S24.

Figure 3a visualizes these structural differences through principal component analysis (PCA) of 2,048-bit, radius-two Morgan fingerprints. PCA was fitted only to a fixed subset of 1,024 training records; the generated structures were then projected onto the same axes. Grey shading shows the density of distinct training structures actually encountered by the selected model. Colored density contours show where distinct canonical repeat units are concentrated under each generation condition; inner contours enclose denser regions. The domain-conditioned libraries occupy a more localized part of the projection than the unconditional library, consistent with the lower internal diversity reported in the SI. Principal components 1 and 2 (PC1 and PC2) combine fingerprint features rather than representing physical properties and together capture only 8.6% of fingerprint variation in the reference subset. The map therefore provides a partial view of structural relationships; visible overlap or separation does not describe all chemical differences between the libraries.

Figure 3b summarizes similarity to encountered training structures by averaging Tanimoto similarity across randomly selected training-generated pairs. The pooled means are 0.160, 0.066, 0.091, and 0.066 for unconditional, electrolyte-only, low-Tg, and combined generation, respectively. The per-seed estimates reported in Table S25 are closely grouped. The lower averages under electrolyte conditioning indicate a shift relative to the broader training collection, rather than greater diversity within the generated library. Close training analogues remain evident in the nearest-neighbor analysis (Figure S7 and Tables S21 and S24). Mean pairwise similarity does not establish novelty, which is assessed earlier separately against the complete eligible training catalog.

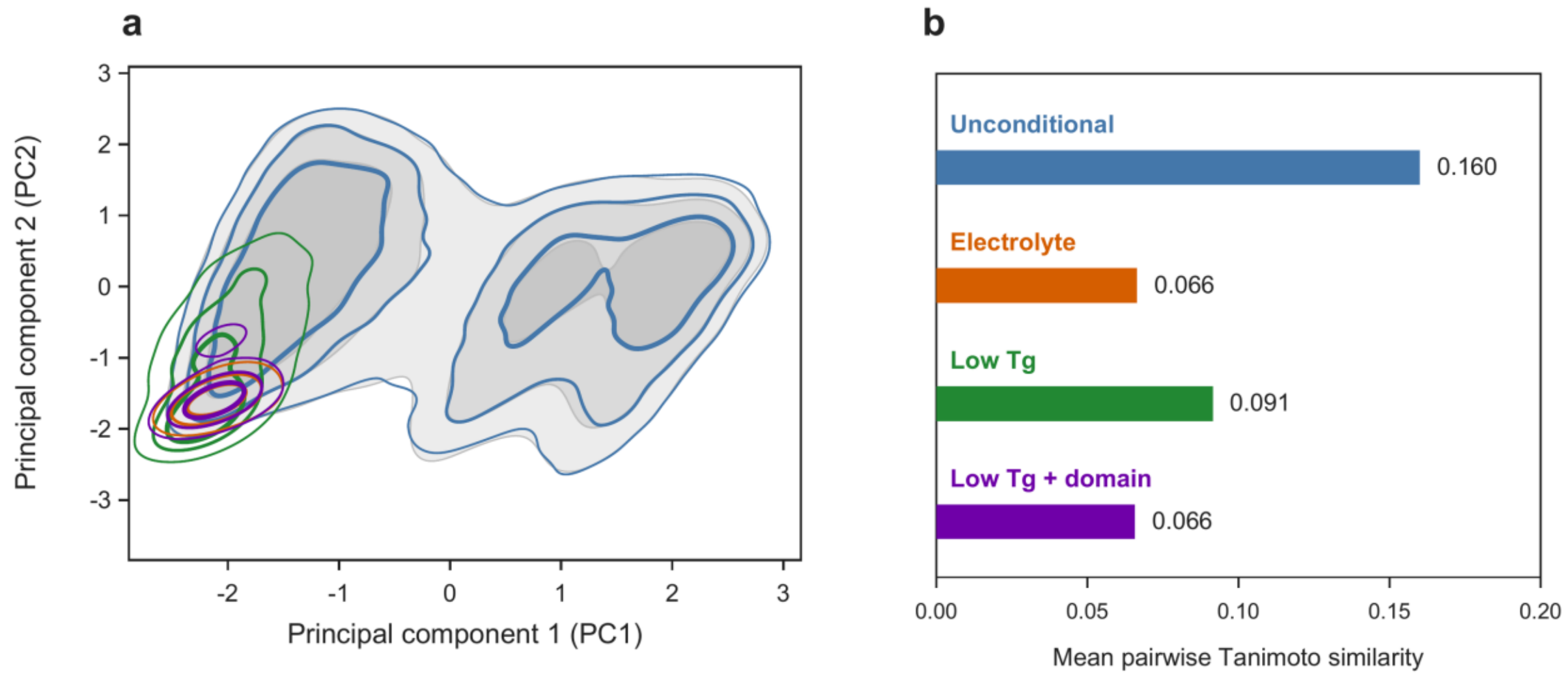

**Figure 3.** Chemical-space distributions and similarity to encountered training structures. (a) Grey shading shows all 34,263,511 distinct canonical repeat-unit structures actually encountered through 43,000 updates, with darker regions indicating higher density. Colored contours show the generated libraries. From inner to outer, contour boundaries enclose 50%, 80%, and 95% of each distribution's estimated density, using a common bandwidth. Projections use a fixed PCA basis fitted to training-set Morgan fingerprints; the first two components explain 8.6% of the fitting-subset variance. (b) Mean pairwise Tanimoto similarity between generated structures and the encountered training reference, using the same fingerprints as panel (a). Colored bars show the pooled mean after canonical deduplication across the three generation seeds. Condition colors are shared with panel (a). Each distinct generated structure and reference structure has equal weight.

## Control of numerical properties

Next, we evaluated numerical requests for the eight properties with available predictors (Figure 4 and Table 2). These tests did not include the electrolyte-domain token. The predicted values shifted toward the requested targets, but the size of the shifts and the prediction errors differed by property. The lower requested Tg in this comparison was 437.06 K; the principal low-Tg libraries instead used 246.11 K. Increasing the requested Tg from 437.06 to 560.54 K increased the mean predictor-estimated Tg by 116.2 K, compared with a requested difference of 123.5 K. The corresponding MAEs were 41.51 and 31.30 K, respectively. Changing the requested Tg shifted the average predictor-estimated Tg, but the mean absolute errors remained between 31 and 42 K.

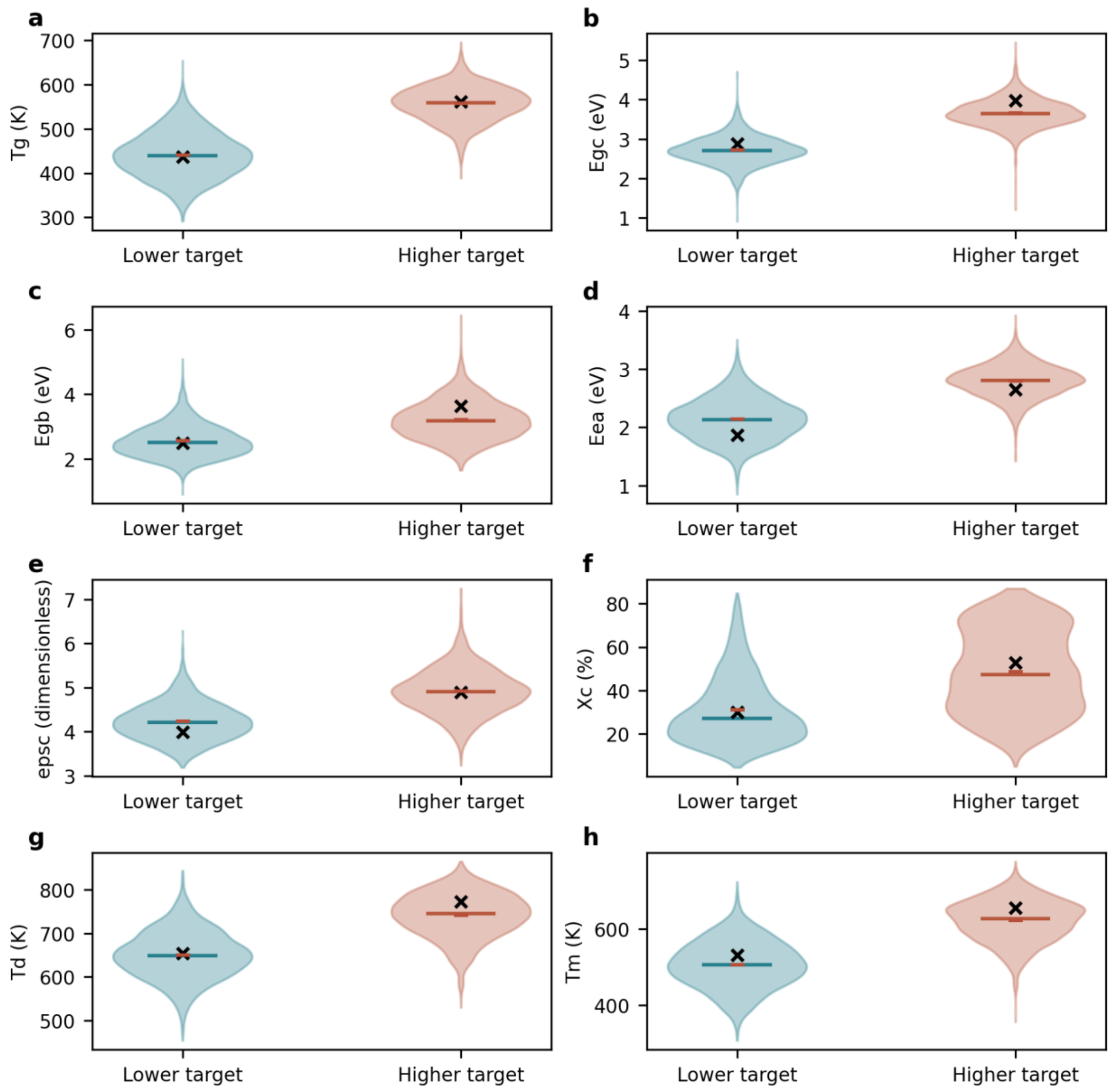

**Figure 4.** Numerical conditioning for (a) glass-transition temperature (Tg), (b) chain electronic band gap (Egc), (c) bulk electronic band gap (Egb), (d) electron affinity (Eea), (e) calculated dielectric constant (epsc), (f) DFT crystallization tendency (Xc), (g) degradation/decomposition temperature (Td), and (h) melting temperature (Tm). These requests omit domain conditioning; scoring coverage is reported in Table S6. Violin plots show predictor-estimated values for lower and higher requested targets; black crosses indicate the targets. The Tg targets are 437.06 and 560.54 K. Properties are assessed for valid structures with available predictions. The dielectric constant is assessed using the static eps_DFT predictor. Long horizontal marks matching the violin colors indicate the medians; short rust-colored marks indicate the means. Violin widths represent density.

**Table 2.** Accuracy of single-property numerical control. MAE is calculated for valid structures with available predictions. Complete scoring coverage, RMSE, signed bias, and useful yields are reported in Table S6. Xc targets are expressed in percent, and Xc errors are expressed in percentage points. The epsc results use the static dielectric predictor.

| Property | Lower target | MAE at lower target | Higher target | MAE at higher target |
|---|---|---|---|---|
| Tg (K) | 437.06 | 41.51 | 560.54 | 31.30 |
| Egc (eV) | 2.88 | 0.31 | 3.97 | 0.38 |
| Egb (eV) | 2.49 | 0.39 | 3.64 | 0.61 |
| Eea (eV) | 1.87 | 0.36 | 2.65 | 0.26 |
| epsc | 3.99 | 0.38 | 4.90 | 0.40 |
| Xc | 29.88 | 12.07 | 52.84 | 16.77 |
| Td (K) | 654.40 | 43.81 | 771.75 | 46.03 |
| Tm (K) | 530.54 | 53.66 | 653.91 | 49.12 |

For the electronic properties, the changes in mean predictions were generally smaller than the changes in the requested targets. The ratio of predicted mean shift to requested shift is 0.87 for Egc, 0.57 for Egb and 0.83 for Eea. Thus, a response in the requested direction does not imply that the full interval is reproduced. The dielectric, Xc and Td ratios are 0.76, 0.76 and 0.78, respectively. For Tm, the ratio is 0.93, with MAEs of 53.66 and 49.12 K. The Xc distributions shift with the requested target but remain broad, indicating substantial variation in predicted Xc among generated structures (Figure 4f).

The assessment uses property models that also supplied some of the conditioning labels. Because some of these predictors also supplied training labels, agreement with a target reflects agreement with those predictors rather than independent verification of the property. Their physical interpretation depends on the accuracy and applicability of the underlying predictors, whose cross-validation results are summarized in Table S5.

## Control of polymer functionality

We next tested whether the generated polymers contained or excluded each requested chemical feature (Figure 5a). Across eleven inclusion requests, the mean fraction of all attempts yielding a valid compliant structure is 93.9%; the mean for exclusion is 97.0%. Inclusion yield ranges from 88.7% for amine to 99.7% for polyphosphazene. Exclusion yield ranges from 94.1% for amide to 99.4% for polyphosphazene. These tests isolate structural control without the electrolyte token (Table S7); the matched evaluation below examines structural and domain conditions together (Figure 6d, 6e).

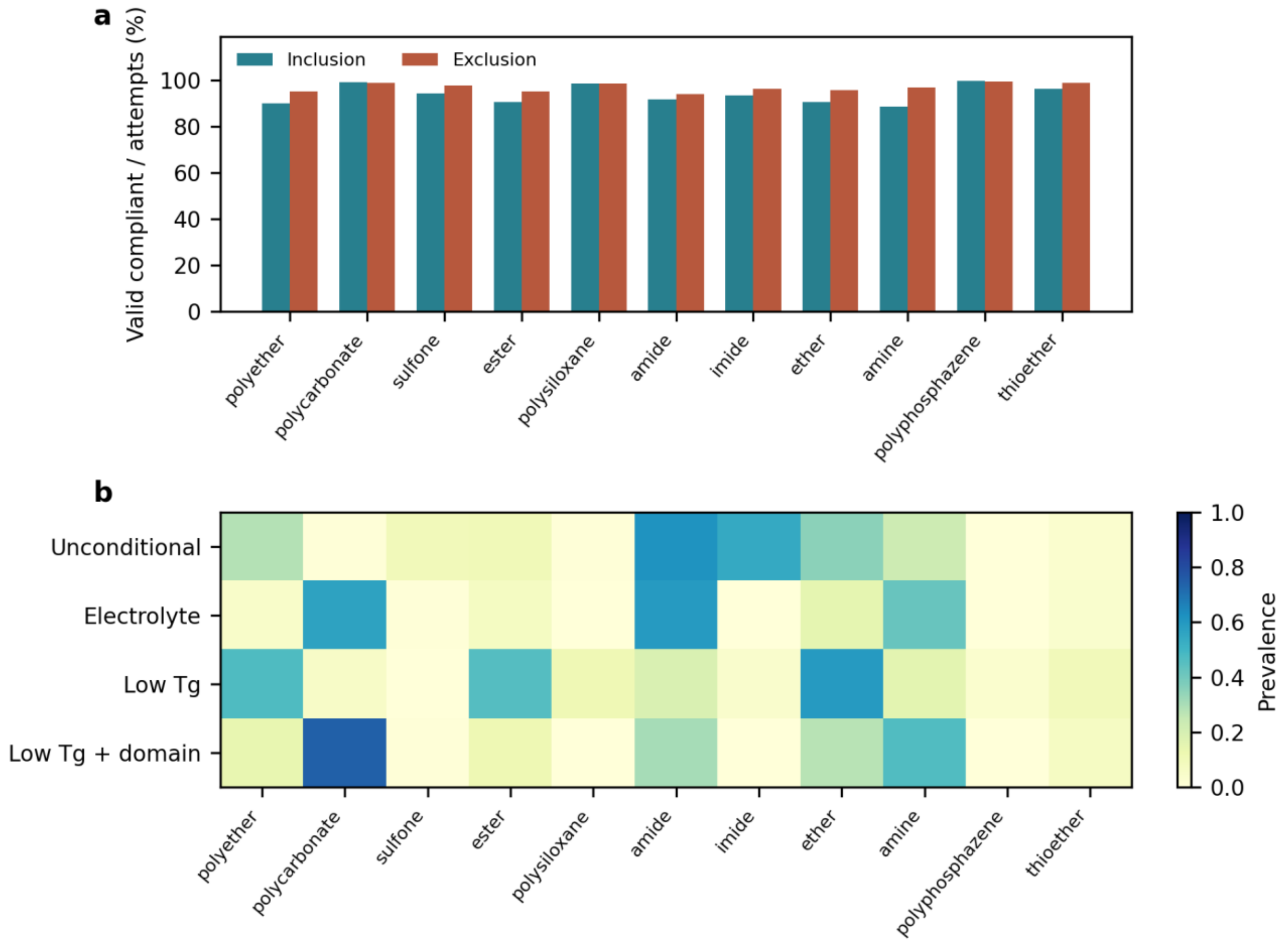

**Figure 5.** Structural control and functional-group composition. (a) Fractions of all attempts yielding valid structures that satisfy one inclusion or exclusion request without domain conditioning. (b) Annotation prevalence in the principal libraries, pooled over three seeds and deduplicated across seeds. Each assessable canonical distinct structure has equal weight; unknown annotations are excluded rather than counted as absent (Tables S21 and S23). Annotations overlap, so fractions do not sum to one. Polyether is a backbone-family annotation, whereas ether denotes the broader motif.

Requesting an ether group tests structural compliance, whereas its prevalence without that request describes the composition favored by the numerical and domain conditions. Electrolyte conditioning increases polycarbonate prevalence from 0.40% to 56.41% and decreases polyether prevalence from 28.35% to 4.85% among distinct structures with known annotations (Figure 5b; Table S23). For low-Tg generation, adding the domain token increases polycarbonate prevalence from 5.65% to 74.56% and decreases ether prevalence from 59.17% to 27.64%. The domain condition therefore changes the favored repeat-unit chemistry even when the requested Tg is unchanged. A repeat can carry several annotations. Note that polyether and ether annotations are not interchangeable.

**Joint conditioning and the support of the training data**

We next examined whether PolyGPT could satisfy a Tg target together with electrolyte-domain conditioning. In the principal libraries, adding the domain condition at 246.11 K slightly reduces predictor-estimated Tg MAE from 26.97 ± 0.15 to 26.07 ± 0.11 K (mean ± SD across seeds; Table S18). This improvement is accompanied by greater duplication. Although useful yield within 50 K of the target rises from 85.89% to 87.25%, distinct useful yield falls from 82.35% to 32.31%. When we also require these structures to be absent from the training catalog, the novel distinct useful yield reduces to 59.90% and 19.71%, respectively. The domain condition therefore results in a slightly lower Tg MAE but a much smaller library of distinct, catalog-unseen candidates that meet the target.

The effect of domain conditioning changes when the requested Tg is increased (Figure 6a). Between targets of 437.06 and 560.54 K, the mean predicted Tg rises by 116.2 K without domain conditioning but by only 29.8 K with it. At 560.54 K, adding the electrolyte condition increases the Tg MAE from 31.30 to 158.93 K. The distinct useful yield within 50 K of the target is 8.9%, calculated relative to all attempts (Table S6). Thus, the domain condition substantially reduces the response to the requested change in Tg over this interval.

The distribution of Tg labels in the electrolyte training collection is consistent with this difference. Among 5,808 electrolyte training groups, 1,728 have a retained $T_g$ label within 25 K of 246.11 K, whereas none has a label within 50 K of 560.54 K. The mean labeled $T_g$ is 291.3 K when groups contribute equally and repeat variants are weighted equally within each group. A low-$T_g$ electrolyte request is therefore supported by many application records, while the high-$T_g$ request combines conditions with little nearby support. The greater training support near the low target is therefore consistent with its better performance. However, these comparisons do not isolate the effect of training-set coverage.

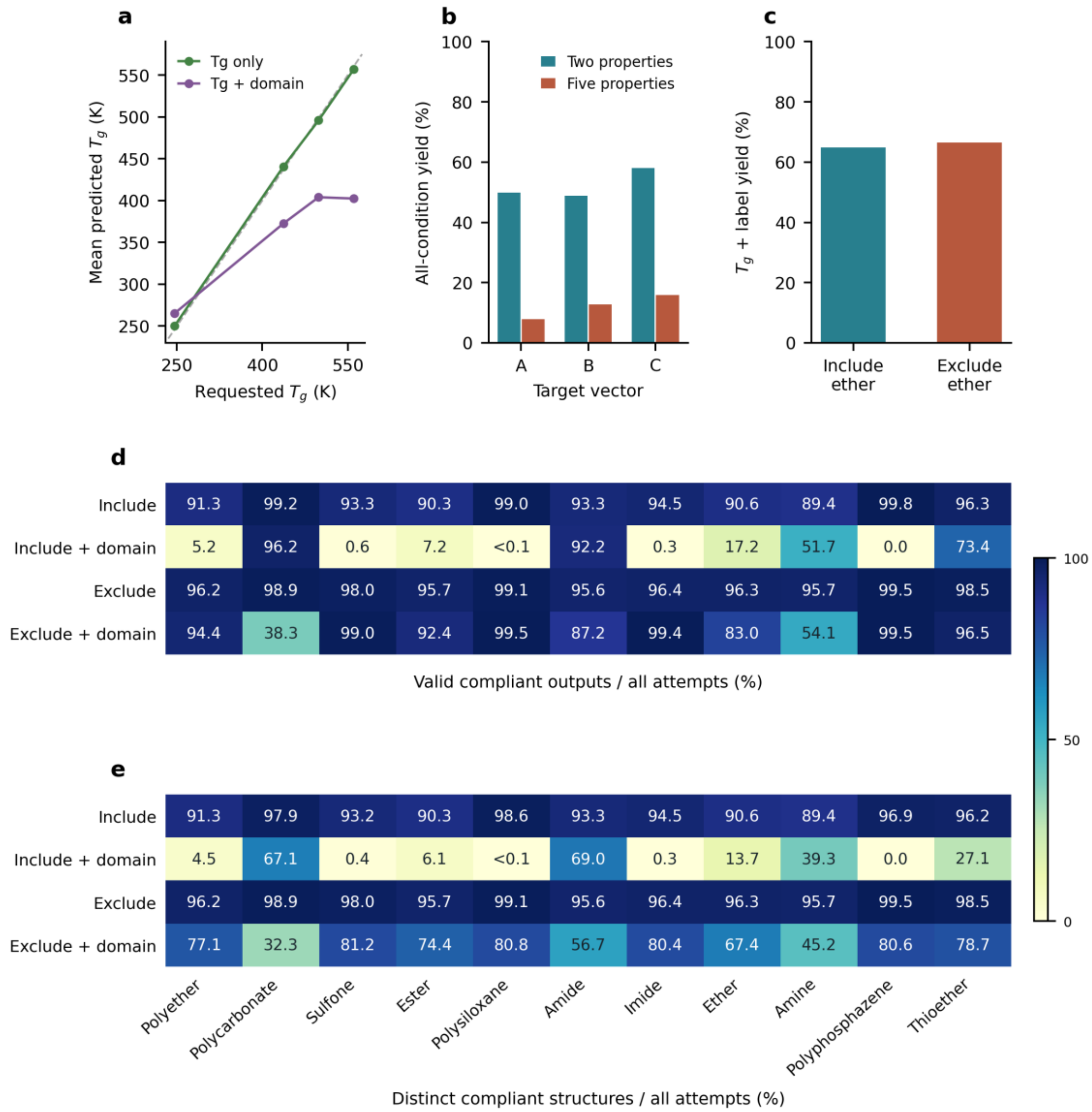

**Figure 6.** Joint conditioning and the yield of compliant structures. (a) Requested versus mean predicted Tg with and without electrolyte conditioning; the dashed line denotes equality between the requested and mean predicted values. (b) Useful yield for two-property (Tg/Egc) and five-property (Tg/Egc/Egb/Eea/Xc) requests using target vectors A-C from Table S8. (c) Useful yield for simultaneous Tg and ether inclusion or exclusion requests at Tg = 498.799 K. Useful outputs satisfy every requested numerical and structural condition. Panels (b, c) omit the domain token. Numerical acceptance bands are defined in Supporting Information Section S1.5. (d) Valid compliant yield and (e) distinct compliant yield for eleven single-label requests, with and without the electrolyte token. No numerical property targets were specified in panels (d, e). Each condition pools three seeds. Both matrices use all attempts as the denominator, with percentages shown in the cells; positive yields below 0.1% are marked <0.1. Unknown annotations are not confirmed label failures and are counted as non-compliant.

When we requested five numerical properties rather than two, fewer generated structures satisfied all requested targets (Figure 6b). Using target vectors drawn from individual training records, $T_g/E_{gc}$ requests give useful yields of 48.8-58.1%. Adding $E_{gb}$, $E_{ea}$, and $X_c$ reduces these yields to 7.8-15.9%, while $T_g$ MAE for the five-property requests is 28.9-38.3 K (Tables S8-S10). Agreement with Tg alone, therefore, conceals much of the loss in simultaneous compliance. $X_c$ MAE is 11.1-16.7 percentage points against a 10-percentage-point acceptance band. The $X_c$ predictor itself has a cross-validation MAE of 12.67 percentage points (Table S5), which limits the physical interpretation of these compliance estimates.

For joint numerical and structural control, requesting $T_g$ = 498.799 K with ether inclusion or exclusion gives useful yields of 64.9% and 66.4%, respectively (Figure 6c). Both requests omit the electrolyte token. For these two requests, the model generated structures that met both the Tg target and the ether inclusion or exclusion requirement.

We then combined each structural request with the electrolyte-domain condition while leaving numerical targets unspecified (Figures 6d, 6e). Across the eleven inclusion requests, the mean valid compliant yield decreases from 94.27% without domain conditioning to 31.27% with it, while the mean distinct compliant yield decreases from 93.83% to 20.69%. For exclusion, the corresponding valid compliant yields are 97.26% and 85.76%, and distinct compliant yields are 97.26% and 68.61%. Polycarbonate exclusion shows a particularly large decrease, from 98.88% to 38.31%. These averages weight each annotation equally; individual requests and sampling variation are reported in Tables S26-S30.

The chemical features differ substantially in how often they occur among electrolyte polymers used for training. Polycarbonate and amide inclusion retain valid compliant yields of 96.18% and 92.20% under domain conditioning; their annotations occur in 3,485 and 3,462 electrolyte training groups, respectively. Only one group supports polyphosphazene, for which domain-conditioned inclusion gives 0.00% compliant yield (Tables S26 and S29). This contrast is consistent with poorer control for sparsely represented combinations, although support counts alone do not establish the cause. Duplication creates another limitation: thioether inclusion with domain conditioning gives 73.36% valid compliant yield but only 27.05% distinct compliant yield. Natural motif prevalence provides a separate check on exclusions: avoiding an already rare feature is a weaker test of control than suppressing a common one (Table S29). These counts describe the available training combinations, not the number of times they were sampled.

Combining conditions can therefore reduce compliance even when each control is effective on its own. These evaluations cover pairwise combinations of control types. Simultaneous numerical, structural, and domain conditioning was not tested for this study.

**Training duration balances control and novelty**

We compared nine checkpoints from one training run at 10,000, 15,000, 20,000, 25,000, 30,000, 35,000, 40,000, 43,000, and 45,000 updates (Figure 7; Table S11). We used the same generation requests, decoding settings, random seeds, and sample sizes for every checkpoint. The primary 43,000-update model was selected by the minimum balanced validation objective, which is 0.2555, compared with 0.2807 at 10,000 and 0.2695 at 20,000 updates. Selection by validation loss does not necessarily maximize the yield of distinct novel candidates.

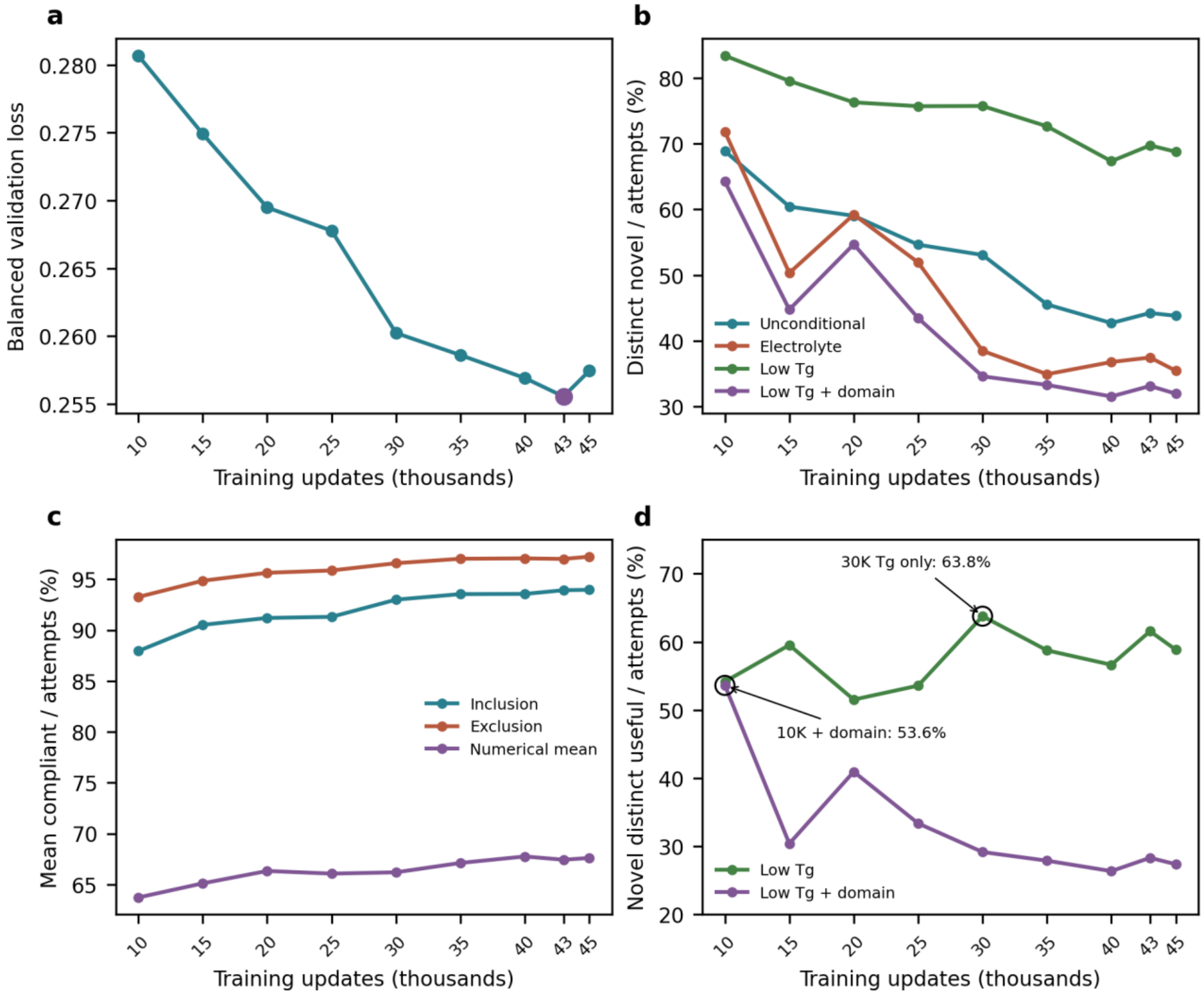

**Figure 7.** Control and novelty across all nine checkpoints. (a) The balanced validation objective, with the selected 43,000-update model marked. (b) Yield of distinct structures absent from the training catalog under the four principal generation conditions. (c) Mean valid compliant yield for eleven inclusion requests and eleven exclusion requests, and mean useful yield for eight properties at their lower/higher targets, each expressed per attempted generation. Numerical compliance is averaged across the eight properties using their respective acceptance tolerances. (d) Novel distinct useful yield within 50 K of 246.11 K, with and without domain conditioning. Panels (b-d) use the checkpoint comparison at equal sample sizes, separately from the principal libraries. All checkpoints come from one training run.

With the same number of generation attempts at each checkpoint, some earlier checkpoints produced more distinct structures absent from the training catalog. Unconditional novel distinct yield decreases from 68.8% at 10,000 updates to 44.2% at 43,000 updates, despite validity increasing from 97.3% to 99.4% and pooled canonical uniqueness remaining 100%. For domain-only generation, novel distinct yield decreases from 71.8% to 37.5%. The sharper distinction appears when novelty and low-Tg compliance must hold together: novel distinct useful domain-conditioned yield is 53.6% at 10,000 updates, 41.0% at 20,000, and 28.3% at 43,000. Tg MAE at 10,000 updates is 28.24 K, 2.03 K above that of the primary model. For this request, the earlier checkpoint therefore has higher novel distinct useful yield with a modest increase in Tg error. At 43,000 updates, novel distinct useful yield is 28.3% in the checkpoint comparison versus 19.71% in the larger principal library, consistent with the greater duplication observed as the sample grows (Figure S4).

For the 30,000-update checkpoint without domain conditioning, the novel distinct useful yield within 50 K of the low target is 63.8%, compared with 61.6% for the primary model. With domain conditioning, its yield is 29.2%, close to the primary model but well below 10,000 updates. The effect of training duration on novel distinct useful yield therefore depends on whether domain conditioning is applied. Across all nine checkpoints, mean inclusion yield rises from 87.9% at 10,000 updates to 93.9% at 43,000, while mean numerical yield rises from 63.7% to 67.4%. Neither measure peaks at the selected checkpoint: the highest observed inclusion mean is at 45,000 updates, and the highest numerical mean is at 40,000.

### Sampling sensitivity

We assessed decoding sensitivity for the selected 43,000-update model. The grid crossed temperatures 0.8, 1.0, and 1.2 with top-k 0, 20, and 50 at top-p 1. Unconditional and low-Tg-plus-electrolyte requests were evaluated at every setting (Figures S2 and S3). Increasing temperature from 1.0 to 1.2 without top-k truncation changes unconditional novel distinct yield from 45.0% to 55.2%, while validity changes from 99.3% to 98.0%. Under the low-Tg/domain request, novel distinct yield changes from 31.3% to 41.8%, and Tg MAE changes from 25.88 to 28.75 K. The corresponding novel distinct useful yields within 50 K are 27.5% and 35.0%. For the low-Tg/domain request, the higher temperature increases novel distinct useful yield despite the larger Tg error. Across the three sampling seeds, Tg MAE (mean and sample SD) is 25.88 ± 0.49 K at temperature 1 and 28.75 ± 0.44 K at temperature 1.2, both without top-k truncation. Top-k truncation at 20 or 50 has a much smaller effect in this grid: relative to untruncated sampling at the same temperature, validity changes by at most 0.13 percentage points and conditional Tg MAE by 0.02 K. This observation is specific to the tested prompts and truncation range.

Changing decoding settings changes which structures are sampled from the selected model; additional training changes the model itself. The principal evaluation used temperature 1, top-k 0, and top-p 1. It is analyzed separately because uniqueness and distinct useful yield depend on the number of generation attempts.

**Target tolerance, library size, and repeat-unit representation**

In the principal libraries, adding domain conditioning also reduces distinct and novel distinct useful yields at the stricter 25K tolerance (Figure 8 and Table 3). Success before removing duplicates is 53.46% without domain and 56.66% with domain, whereas distinct useful yield is 51.40% and 18.98%; novel distinct useful yield is 38.07% and 11.72%. Although domain conditioning slightly lowers Tg MAE from 26.97 to 26.07 K, it increases the positive bias from +5.10 to +19.31 K (Table S18). Its lower MAE should therefore not be interpreted as closer agreement between the mean predicted Tg and the requested value. The decrease in distinct useful yield is observed at both tolerances.

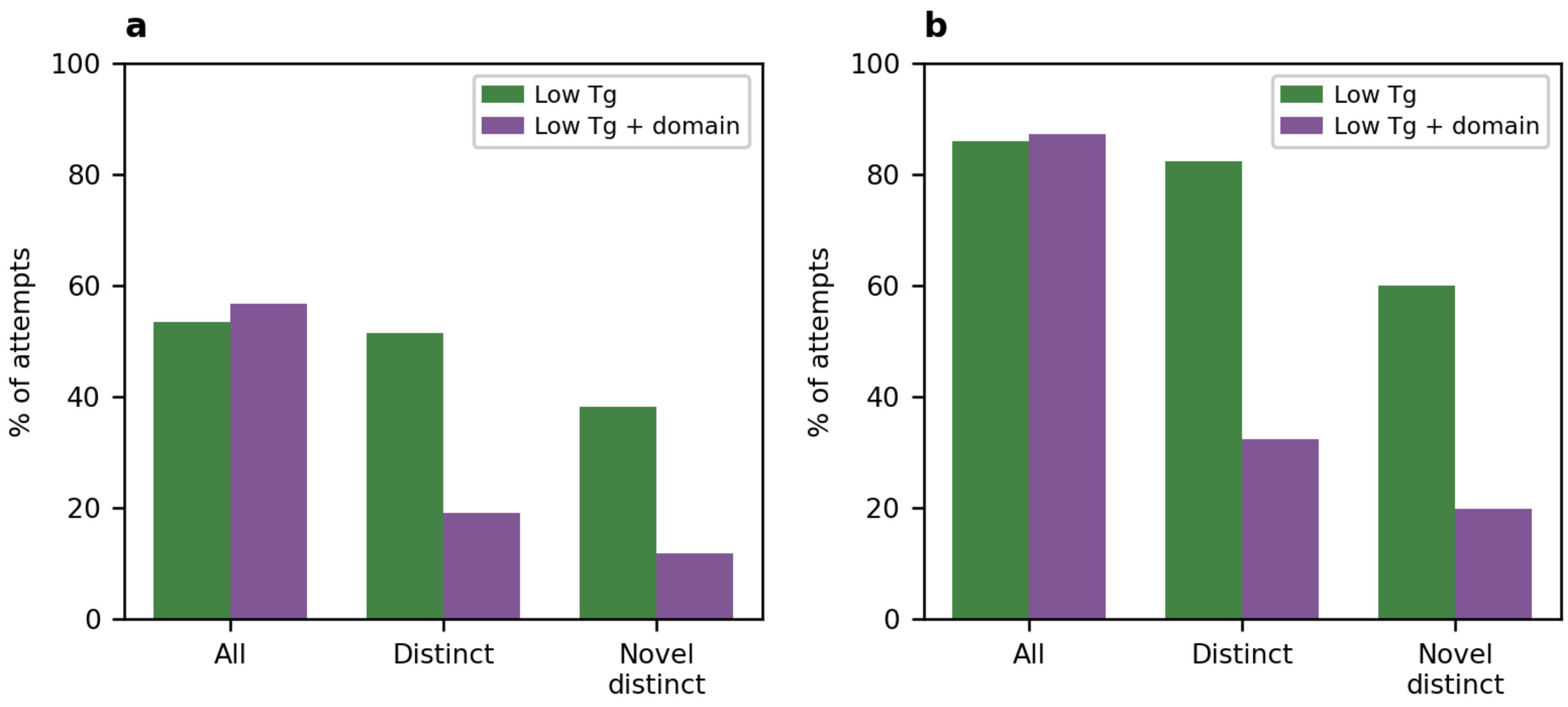

**Figure 8.** Tolerance sensitivity in the principal low-Tg libraries within (a) ±25 K and (b) ±50 K of the 246.11 K target. All successes count every valid scored draw within the stated Tg band; distinct successes remove canonical duplicates; novel distinct successes also require verified absence from the complete eligible training catalog.

**Table 3.** Counts and yields under two Tg tolerances. Every yield uses all attempted generations per condition, including invalids and duplicates.

| Condition | Band (K) | All successes | Distinct successes | Novel distinct successes | Distinct yield (%) | Novel yield (%) |
|---|---|---|---|---|---|---|
| Low Tg | 25 | 16,037 | 15,419 | 11,422 | 51.40 | 38.07 |
| Low Tg | 50 | 25,768 | 24,705 | 17,970 | 82.35 | 59.90 |
| Low Tg + domain | 25 | 16,998 | 5,694 | 3,515 | 18.98 | 11.72 |
| Low Tg + domain | 50 | 26,174 | 9,693 | 5,912 | 32.31 | 19.71 |

We analyzed nested prefixes, meaning successively longer portions of the same generation sequence from each random seed. With electrolyte-domain requests, adding more attempts produced fewer new distinct repeat units than unconditional or Tg-only generation (Figure S4). For the combined Tg/domain request, mean within-seed uniqueness was 58.02%: the number of distinct valid canonical repeat units was divided by the number of valid outputs for each seed, and the three percentages were then averaged. Pooling all three seeds, removing duplicates across the combined library and dividing the remaining distinct count by all valid outputs from the three seeds gave 39.08% uniqueness, revealing structures shared between seeds. We also applied conservative periodic grouping, an approximate check that groups some alternative repeat-unit descriptions of a repeating chain, such as a short repeat and two consecutive copies of it. Without a Tg filter, this check reduced the domain-only count from 15,458 distinct canonical repeat units (different finite repeat-unit structures) to 12,881 groups and the low-Tg/domain count from 11,664 to 9,580 groups (Table S19).

Figure S5 illustrates the sensitivity of property estimates to repeat-unit representation. The two representations assigned to conservative group P1 receive Tg estimates of 326.4 and 348.9 K, while those in group P2 receive 262.3 and 244.5 K. These differences of 22.5 and 17.8 K show that the predictor remains sensitive to the alternative representations grouped by this procedure.

Figure 9 shows the first valid output from each condition and sampling seed. The selection rule uses neither predicted Tg, novelty nor visual appearance. The displayed unconditional structures contain several aromatic rings, whereas the electrolyte-conditioned examples have smaller, predominantly aliphatic repeat units. The low-Tg-only examples contain extended aliphatic segments. These structures illustrate the size and compositional differences quantified across the complete libraries in Figures 3 and 5b and Tables S21 and S23. Structural validity alone does not establish synthesis feasibility or thermodynamic stability.

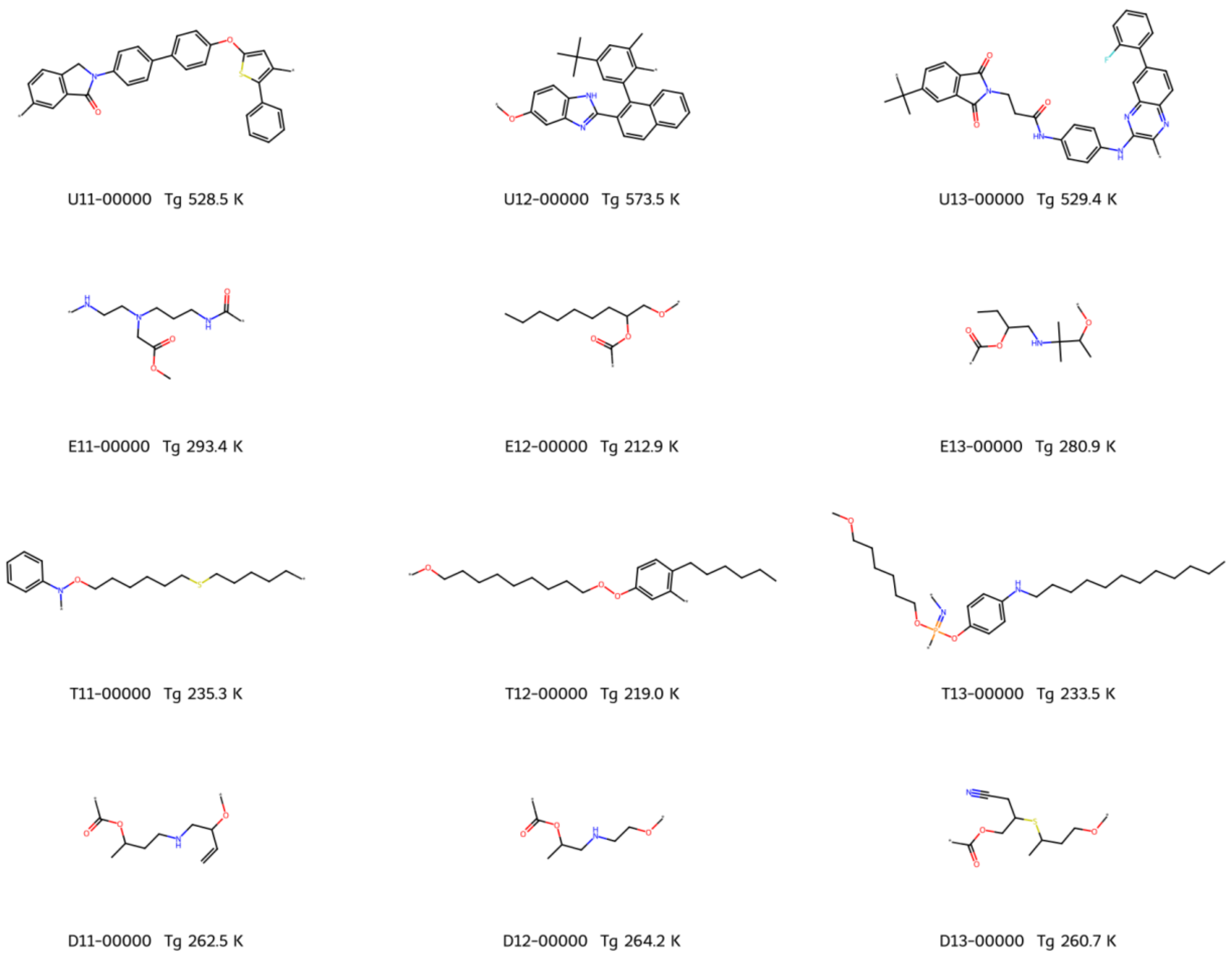

**Figure 9.** First valid repeat unit generated under each condition and random seed. Rows correspond to unconditional (U), electrolyte-only (E), low-$T_g$-only (T), and low-$T_g$-plus-domain (D) generation. Columns correspond to seeds ending in 11, 12, and 13. Identifiers include the attempt number, starting at zero. Displayed $T_g$ values are predictor estimates, not measurements.

## Conclusions

The PolyGPT evaluations show that lower predictor-estimated target error can accompany a smaller set of distinct candidates for further study. Electrolyte conditioning narrowed the generated chemical distribution and increased duplication, so high structural validity and low property error were insufficient to describe its value for candidate discovery. For equal numbers of generation attempts, we therefore considered both the number of distinct structures meeting each request and the number absent from the training catalog.

These findings also distinguish selection by validation loss from selection for a particular generation task. The training duration and sampling settings that favor novel distinct useful yield depend on the requested conditions and the acceptable property error. Weak inclusion control for several sparsely represented domain-structure combinations suggests that training-set coverage deserves closer study, although its contribution cannot be isolated from the present comparisons.

Independent assessment of generated candidates is the next step toward establishing material performance. The numerical evaluations here measure consistency with property predictors, some of which supplied the conditioning labels. Calculations or measurements independent of those predictors are needed to test the property values and evaluate conductivity and stability. Consistent treatment of periodically equivalent repeat units would improve both candidate counting and property comparison. Evaluations across independent training runs and simultaneous numerical, structural, and domain requests would test the reproducibility of the observed tradeoffs and extend the scope of the model assessment.

## Methods

### Data and polymer representation

We used 102,520,074 eligible hypothetical repeat-unit records as the foundation dataset from the polymer collection associated with polyBERT and polyOne [5, 17] and an electrolyte-associated collection of 5,935 conservative polymer groups from high-throughput molecular-dynamics (HTP-MD) [2, 18, 19] and experimental sources [1, 20-23]. Training, validation, and test sets were partitioned by polymer group. The electrolyte token denotes source association rather than a conductivity class. Preparation criteria, source overlap, and partition sizes are given in Supporting Information Section S1.1.

Numerical conditioning covered band gaps, electron affinity, dielectric constant, crystallization tendency, and thermal properties. The nine property channels, normalization statistics, and label provenance are defined in Table S2 and Supporting Information Sections S1.2 and S1.4. Structural conditioning specified inclusion or exclusion of one of eleven potentially overlapping annotations (Table S3).

**Conditional language model**

PolyGPT is a conditional, decoder-only autoregressive Transformer (Table 4). Each numerical target is paired with a presence indicator that distinguishes an unspecified property from a specified zero. A linear projection maps the nine standardized values and nine indicators to one conditioning vector. Optional tokens encode electrolyte association and a structural condition. The network predicts the polymer sequence one token at a time from the preceding tokens and supplied conditions. Unconditional generation leaves all numerical properties and optional controls unspecified. Tokenization and prompt construction are described in Supporting Information Section S1.3.

**Model training**

Training minimized the negative log likelihood of the polymer sequence, including the end-of-sequence token. The objective balanced unconditional, numerical, structural, and combined numerical-structural generation, with equal sampling across modes and weighting across structural requests. No property-prediction term or conductivity reward was used. Loss weighting, electrolyte sampling and exposure limits, and optimization settings are specified in Supporting Information Section S1.3 and Table S4.

The 43,000-update checkpoint was selected by the lowest balanced validation loss, before comparing generated structures across checkpoints. Training ended at 49,000 updates. The nine checkpoints compared in this study were saved at different points in the same training run.

**Table 4.** Model architecture.

| Parameter | Setting |
|---|---|
| Trainable parameters | 38,301,184 |
| Transformer blocks | 12 |
| Hidden width | 512 |
| Attention heads | 8 |
| Feed-forward width | 2,048 |
| Embeddings and weights | Learned position and type embeddings; tied input and output weights |
| Vocabulary size | 388 tokens |

### Generation and conditioning experiments

Generation evaluations used three sampling seeds and the designs and sample sizes in Table 5. We analyzed each evaluation using its own sample sizes and requests; the attempts were not pooled across evaluation designs. Invalid and duplicate outputs remained in the attempt counts and were not replaced.

**Table 5.** Generation evaluations and sample sizes.

| Evaluation | Design | Generation sample size | Total attempts |
|---|---|---|---|
| Principal generation | Four conditions at 43,000 updates: unconditional, electrolyte-domain, low $T_g$, and low $T_g$ + domain | 10,000 per seed and condition; 30,000 pooled per condition | 120,000 |
| Checkpoint comparison | The same 48 core conditions at all nine checkpoints, plus eight joint conditions at seven checkpoints | 1,024 per seed, condition, and checkpoint; 3,072 pooled per condition and checkpoint | 1,499,136 |
| Library accumulation | Nested prefixes of the principal libraries | 1,000, 2,000, 5,000, and 10,000 per seed and condition | Reuses principal samples |

| Evaluation | Design | Generation sample size | Total attempts |
|---|---|---|---|
| Structural/domain control | Eleven labels × inclusion/exclusion × domain present/absent = 44 conditions at 43,000 updates; no numerical targets | 2,048 per seed and condition; 6,144 pooled per condition | 270,336 |
| Sampling sensitivity | At 43,000 updates: unconditional and $T_g$ + domain; temperatures 0.8, 1.0, and 1.2 crossed with top-k 0, 20, and 50; top-p = 1 | 1,024 per seed, condition, and setting | 55,296 |

The nine checkpoints were evaluated at 10,000, 15,000, 20,000, 25,000, 30,000, 35,000, 40,000, 43,000, and 45,000 updates. Joint conditions were evaluated at all of these except 10,000 and 15,000 updates.

The principal and sampling-sensitivity low-Tg conditions used a target of 246.11 K; other numerical and structural conditions were unspecified. Chemical-space and composition analyses used the principal libraries. Decoding settings, random seeds, and computational details are given in Supporting Information Section S1.5.

Combined numerical tests used two- and five-property targets taken from training records; additional tests combined the foundation mean Tg with ether inclusion or exclusion (Table S8 and Supporting Information Section S2.2). Comparisons of numerical conditioning with and without the electrolyte token were restricted to $T_g$.

## Generation quality and property assessment

Generation metrics and their denominators are defined in Table 6; validity criteria are given in Supporting Information Section S1.5. We counted distinct and novel distinct useful structures using canonical repeat-unit representations. Conservative periodic grouping provides a separate assessment of sensitivity to repeat-unit representation (Supporting Information Sections S1.6 and S4.3). Training-catalog novelty is measured against the complete eligible foundation training catalog and electrolyte variants in the training partition, including records not encountered by the evaluated checkpoint. Outputs with unresolved membership are not counted as novel. All criteria for a joint yield must pass in the same candidate [15, 16].

**Table 6.** Definitions of generation metrics.

| Metric | Counted outputs | Denominator |
|---|---|---|
| Structural validity | Outputs satisfying all repeat-unit validity criteria | All attempts |
| Canonical uniqueness | Distinct valid canonical repeat-unit strings | Valid outputs |
| Distinct yield | Distinct valid canonical repeat-unit strings | All attempts |
| Novel distinct yield | Distinct valid canonical identities with verified absence from the training catalog | All attempts |
| Useful yield | Valid, scorable outputs meeting every requested numerical acceptance band and any simultaneous structural request | All attempts |
| Distinct useful yield | Distinct canonical identities among useful outputs | All attempts |
| Novel distinct useful yield | Distinct useful canonical identities with verified absence from the complete eligible training catalog | All attempts |
| Valid compliant yield | Valid outputs with assessable annotations satisfying the requested structural inclusion or exclusion | All attempts |

Eight fixed TransPolymer-based predictors [4] assessed numerical control. When a generated structure matched a reference polymer group, we predicted its properties using only models trained without that group. For other structures that could be scored, we averaged predictions from five models. We report scoring coverage, mean absolute error, root-mean-square error, signed bias, and useful yield. Predictor training and scoring rules, including numerical acceptance bands and exclusions, are specified in Supporting Information Sections S1.4 and S1.5. Because these predictors also supplied some conditioning labels, the results quantify agreement with the property models rather than independently measured material properties.

The property predictors used DFT-derived electronic, crystallization-tendency and static-dielectric data [24], the $T_g$ compilation [25], and merged Tm and Td datasets [26–30]. Data sources and preparation are described in Supporting Information Section S1.4.

Chemical-space analyses weighted each distinct valid canonical structure equally within a condition. Diversity and similarity were assessed using Morgan fingerprints and Tanimoto similarity; Figure 3b reports mean pairwise similarity following Yang et al. [2]. The reference contains structures encountered through 43,000 updates, whereas novelty uses the complete eligible training catalog. PCA projection, density estimation, pair sampling, nearest-neighbor comparisons, and functional-group analyses are detailed in Supporting Information Section S6 and Tables S21-S25.

We calculated means and sample standard deviations across three generation seeds to describe how results varied when sampling from the same trained model. For the principal libraries, results were computed per seed and after pooling. Uncertainty calculations and example selection follow Supporting Information Section S1.5; the reported intervals exclude property-predictor uncertainty.

**Supporting information**

The Supporting Information reports predictor performance; numerical, structural, and combined-conditioning evaluations; comparisons of nine training checkpoints; sampling uncertainty;

tolerance and accumulation analyses; canonical and conservative periodic counts; examples of generated structures; and the held-out sequence-likelihood evaluation.

## Data and code availability

Dataset sources are identified in Supporting Information Sections S1.1 and S1.4. Additional data and code, including the full processed datasets, model and predictor weights, retained training checkpoints and evaluation records not supplied in the supplementary files, are available from the corresponding author upon reasonable request. Third-party data remain subject to their original access and redistribution conditions.

## Declaration of generative AI and AI-assisted technologies

The authors used OpenAI ChatGPT and Codex to assist with manuscript drafting and revision, and literature searches.

## References

[1] Bradford, Gabriel; Lopez, Jeffrey; Ruza, Jurgis; Stolberg, Michael A.; Osterude, Richard; Johnson, Jeremiah A.; Gomez-Bombarelli, Rafael; Shao-Horn, Yang. Chemistry-Informed Machine Learning for Polymer Electrolyte Discovery. ACS Central Science 2023, 9 (2), 206-216. https://doi.org/10.1021/acscentsci.2c01123

[2] Yang, Zhenze; Ye, Weike; Lei, Xiangyun; Schweigert, Daniel; Kwon, Ha-Kyung; Khajeh, Arash. De novo design of polymer electrolytes using GPT-based and diffusion-based generative models. npj Computational Materials 2024, 10, 296. https://doi.org/10.1038/s41524-024-01470-9

[3] Khajeh, Arash; Lei, Xiangyun; Ye, Weike; Yang, Zhenze; Hung, Linda; Schweigert, Daniel; Kwon, Ha-Kyung. A materials discovery framework based on conditional generative models applied to the design of polymer electrolytes. Digital Discovery 2025, 4, 11-20. https://doi.org/10.1039/D4DD00293H

[4] Xu, Changwen; Wang, Yuyang; Barati Farimani, Amir. TransPolymer: a Transformer-based language model for polymer property predictions. npj Computational Materials 2023, 9, 64. https://doi.org/10.1038/s41524-023-01016-5

[5] Kuenneth, Christopher; Ramprasad, Rampi. polyBERT: a chemical language model to enable fully machine-driven ultrafast polymer informatics. Nature Communications 2023, 14, 4099. https://doi.org/10.1038/s41467-023-39868-6

[6] Qiu, Haoke; Liu, Lunyang; Qiu, Xuepeng; Dai, Xuemin; Ji, Xiangling; Sun, Zhao-Yan. PolyNC: a natural and chemical language model for the prediction of unified polymer properties. Chemical Science 2024, 15 (2), 534-544. https://doi.org/10.1039/D3SC05079C

[7] Batra, Rohit; Dai, Hanjun; Huan, Tran Doan; Chen, Lihua; Kim, Chiho; Gutekunst, Will R.; Song, Le; Ramprasad, Rampi. Polymers for Extreme Conditions Designed Using Syntax-Directed Variational Autoencoders. Chemistry of Materials 2020, 32 (24), 10489-10500. https://doi.org/10.1021/acs.chemmater.0c03332

[8] Liu, Di-Fan; Zhang, Yong-Xin; Dong, Wen-Zhuo; Feng, Qi-Kun; Zhong, Shao-Long; Dang, Zhi-Min. High-Temperature Polymer Dielectrics Designed Using an Invertible Molecular Graph Generative Model. Journal of Chemical Information and Modeling 2023, 63 (24), 7669-7675. https://doi.org/10.1021/acs.jcim.3c01572

[9] Gurnani, Rishi; Kamal, Deepak; Tran, Huan; Sahu, Harikrishna; Scharm, Kenny; Ashraf, Usman; Ramprasad, Rampi. polyG2G: A Novel Machine Learning Algorithm Applied to the Generative Design of Polymer Dielectrics. Chemistry of Materials 2021, 33 (17), 7008-7016. https://doi.org/10.1021/acs.chemmater.1c02061

[10] Qiu, Haoke; Sun, Zhao-Yan. On-demand reverse design of polymers with PolyTAO. npj Computational Materials 2024, 10, 273. https://doi.org/10.1038/s41524-024-01466-5

[11] Savit, Anagha; Sahu, Harikrishna; Shukla, Shivank S.; Xiong, Wei; Ramprasad, Rampi. polyBART: A Chemical Linguist for Polymer Property Prediction and Generative Design. In Findings of the Association for Computational Linguistics: EMNLP 2025; Association for Computational Linguistics, 2025; pp 12104-12119. https://doi.org/10.18653/v1/2025.findings-emnlp.647

[12] Sahu, Harikrishna; Xiong, Wei; Savit, Anagha; Shukla, Shivank S.; Ramprasad, Rampi. POLYT5: an encoder-decoder foundation chemical language model for generative polymer design. npj Artificial Intelligence 2026, 2, 30. https://doi.org/10.1038/s44387-026-00087-1

[13] Pyle, Charlie; Gadari, Adarsh; Figg, C. Adrian; Jia, Zhenquan; Li, Yaohang; Zhu, Chunjiang. PolymerGPT: Multi-property Optimization with a Decoder-Based GPT Model for Generative Polymer Design. arXiv preprint 2026, arXiv:2608.01431. https://arxiv.org/abs/2608.01431

[14] Dobberstein, Niklas; Maass, Astrid; Hamaekers, Jan. Llamol: a dynamic multi-conditional generative transformer for de novo molecular design. Journal of Cheminformatics 2024, 16, 73. https://doi.org/10.1186/s13321-024-00863-8

[15] Polykovskiy, Daniil; Zhebrak, Alexander; Sanchez-Lengeling, Benjamin; Golovanov, Sergey; Tatanov, Oktai; Belyaev, Stanislav; Kurbanov, Rauf; Artamonov, Aleksey; Aladinskiy, Vladimir; Veselov, Mark; Kadurin, Artur; Johansson, Simon; Chen, Hongming; Nikolenko, Sergey; Aspuru-Guzik, Alán; Zhavoronkov, Alex. Molecular Sets (MOSES): A Benchmarking Platform for Molecular Generation Models. Frontiers in Pharmacology 2020, 11, 565644. https://doi.org/10.3389/fphar.2020.565644

[16] Brown, Nathan; Fiscato, Marco; Segler, Marwin H. S.; Vaucher, Alain C. GuacaMol: Benchmarking Models for de Novo Molecular Design. Journal of Chemical Information and Modeling 2019, 59 (3), 1096-1108. https://doi.org/10.1021/acs.jcim.8b00839

[17] Kuenneth, Christopher; Ramprasad, Rampi. polyOne Data Set - 100 million hypothetical polymers including 29 properties. Zenodo 2022, version 1. https://doi.org/10.5281/zenodo.7124188

[18] Xie, Tian; France-Lanord, Arthur; Wang, Yanming; Lopez, Jeffrey; Stolberg, Michael A.; Hill, Megan; Leverick, Graham Michael; Gomez-Bombarelli, Rafael; Johnson, Jeremiah A.; Shao-Horn, Yang; Grossman, Jeffrey C. Accelerating amorphous polymer electrolyte screening by learning to reduce errors in molecular dynamics simulated properties. Nature Communications 2022, 13, 3415. https://doi.org/10.1038/s41467-022-30994-1

[19] Xie, Tian; Kwon, Ha-Kyung; Schweigert, Daniel; Gong, Sheng; France-Lanord, Arthur; Khajeh, Arash; Crabb, Emily; Puzon, Michael; Fajardo, Chris; Powelson, Will; Shao-Horn, Yang; Grossman, Jeffrey C. A cloud platform for sharing and automated analysis of raw data from high throughput polymer MD simulations. APL Machine Learning 2023, 1 (4), 046108. https://doi.org/10.1063/5.0160937

[20] Schauser, Nicole S.; Kliegle, Gabrielle A.; Cooke, Piper; Segalman, Rachel A.; Seshadri, Ram. Database Creation, Visualization, and Statistical Learning for Polymer Li+-Electrolyte Design. Chemistry of Materials 2021, 33 (13), 4863-4876. https://doi.org/10.1021/acs.chemmater.0c04767

[21] Hatakeyama-Sato, Kan; Tezuka, Toshiki; Umeki, Momoka; Oyaizu, Kenichi. AI-Assisted Exploration of Superionic Glass-Type Li+ Conductors with Aromatic Structures. Journal of the American Chemical Society 2020, 142 (7), 3301-3305. https://doi.org/10.1021/jacs.9b11442

[22] Ruza, Jurgis; Stolberg, Michael. Experimental data for "Autonomous Discovery of Polymer Electrolyte Formulations with Warm-start Batch Bayesian Optimization". Zenodo 2025, version v1. https://doi.org/10.5281/zenodo.16741235

[23] Stolberg, Michael A.; Lopez, Jeffrey; Cawthern, Sawyer D.; Herzog-Arbeitman, Abraham; Kwon, Ha-Kyung; Schweigert, Daniel; Anapolosky, Abraham; Storey, Brian D.; Johnson, Jeremiah A.; Shao-Horn, Yang. A data-driven platform for automated characterization of polymer electrolytes. Matter 2025, 8 (6), 102129. https://doi.org/10.1016/j.matt.2025.102129

[24] Kuenneth, Christopher; Rajan, Arunkumar Chitteth; Tran, Huan; Chen, Lihua; Kim, Chiho; Ramprasad, Rampi. Polymer informatics with multi-task learning. Patterns 2021, 2 (4), 100238. https://doi.org/10.1016/j.patter.2021.100238

[25] linyeping. Extra dataset with SMILES,$T_g$,PID,Polimers Class. Kaggle dataset. https://www.kaggle.com/datasets/linyeping/extra-dataset-with-smilestgpidpolimers-class

[26] Tang, Hao; Yue, Tianle; Li, Ying. Assessing Uncertainty in Machine Learning for Polymer Property Prediction: A Benchmark Study. Journal of Chemical Information and Modeling 2025, 65 (13), 6585-6598. https://doi.org/10.1021/acs.jcim.5c00550

[27] Jupiterj. polyLearn. GitHub repository, revision e4c20cb59e45. https://github.com/Jupiterj/polyLearn

[28] WangGroupFDU. Openpoly_benchmark. GitHub repository, revision ccaa66900747. https://github.com/WangGroupFDU/Openpoly_benchmark

[29] BoranLilike. PolyFKAN: Integrating Fourier-KAN Layers into Graph Neural Networks for Polymer Property Prediction. GitHub repository, revision 83dcafb32126. https://github.com/BoranLilike/PolyFKAN

[30] Otsuka, Shingo; et al. PoLyInfo: Polymer Database for Polymeric Materials Design. In 2011 International Conference on Emerging Intelligent Data and Web Technologies; 2011; pp 22-29. https://doi.org/10.1109/EIDWT.2011.13

# Supporting Information

PolyGPT: A Conditional Generative Transformer for Polymer Electrolyte Design

Tanmay Sarkar Akash, Siddhartha Das*
University of Maryland
*Corresponding author

## S1 Supplementary methods

### S1.1 Data sources and preparation

The foundation release associated with polyBERT [1] and polyOne [2] contained 102,702,346 prepared repeat-unit records. The electrolyte sources contributed 26,685 entries corresponding to 6,183 exact repeat-unit variants. Source entries can overlap, and alternative repeat-unit representations can describe the same polymer. Source-specific entry counts are given in Table S1.

Eligible records contain a supported, connected repeat unit with two attachment sites, a conservative group identifier, lossless tokenization, and at most 397 chemical tokens. Unresolved annotations remain unknown; annotation failure alone does not exclude a record that can still be grouped. Unsupported mixtures are excluded. Generator training used no data augmentation.

Foundation and electrolyte records were assigned to common, deterministic partitions by polymer group: approximately 98% for training, 1% for validation and 1% for testing. The assignment used a fixed hash of the group identity. The resulting training, validation and test sets contained 100,469,281, 1,026,445 and 1,024,348 foundation records; 5,808, 70 and 57 electrolyte groups; and 6,054, 71 and 58 electrolyte variants, respectively.

Electrolyte records required a resolved repeat unit and an explicit association with lithium electrolytes, without a conductivity threshold. HTP-MD entries are simulated; the other retained sources are experimental. Conductivity, temperature and salt concentration were not generation targets.

**Table S1.** Electrolyte data sources. Counts are source entries and include overlap, notably between the raw and cleaned Bradford collections; they are not counts of independent polymers or measurements.

| Source | Entries | Source version or identifier |
|---|---|---|
| Bradford raw [3] | 9,159 | Chem-prop-pred, revision 39b0c16 |
| Schauser [4] | 1,338 | PolymerElectrolyte, revision 5935b20 |
| Hatakeyama [5] | 80 | ion-predictor, revision a58c08c |
| HTP-MD [6-8] | 6,024 | PolyGen, revision 5584a57 |
| PCL [9] | 711 | Zenodo record 16741235 |
| PEO [10] | 192 | Data S1, DOI: 10.1016/j.matt.2025.102129 |
| ChemArr cleaned [3] | 9,181 | Chem-prop-pred, revision 39b0c16 |

### S1.2 Numerical and structural conditioning

Numerical channels follow the order listed in Table S2. For a property that was not requested, we set both its standardized input and its presence indicator to zero. The model therefore distinguished an unspecified property from a requested value of zero.

In the source dictionary, $X_c$ and $X_e$ denote DFT-derived and experimentally referenced crystallization tendency, respectively, in percent (Table 1 in Ref. 1). Foundation labels are model estimates from polyOne [2], including properties whose reference data are experimental.

We copied property labels to 67 electrolyte variants only when their structures and chemical-token sequences matched exactly; we retained the original labels. TransPolymer predictors [11] supplied missing values for eight channels across all 6,183 variants. $X_e$ remained present for 67 and absent for 6,116 variants. The static dielectric predictor eps-DFT supplied and assessed epsc, but its calibration to the foundation-label predictor and to frequency-specific experimental measurements was not established.

**Table S2.** Numerical channels and training-set standardization. Each channel has 100,469,281 observed training values. SD is the population standard deviation. No logarithmic transformation is used.

| Channel | Endpoint | Unit | Mean | SD |
|---|---|---|---|---|
| $E_{gc}$ | Chain electronic band gap | eV | 3.425330 | 0.540986 |
| $E_{gb}$ | Bulk electronic band gap | eV | 3.063908 | 0.575652 |
| $E_{ea}$ | Electron affinity | eV | 2.257784 | 0.392723 |
| epsc | Calculated dielectric constant | dimensionless | 4.446163 | 0.452163 |
| $X_c$ | DFT crystallization tendency | % | 41.357402 | 11.477954 |
| $X_e$ | Experimental crystallization tendency | % | 38.259706 | 9.249860 |
| $T_d$ | Degradation/decomposition temperature | K | 713.075792 | 58.677096 |
| $T_g$ | Glass-transition temperature | K | 498.798707 | 61.742178 |
| $T_m$ | Melting temperature | K | 592.226863 | 61.685066 |

Available values are centered on the foundation-training mean and divided by its population standard deviation. We applied the same foundation-training means and standard deviations when scaling the electrolyte data, including the values predicted for missing labels. In numerical conditioning, the number of requested properties is sampled uniformly from one to the number available; that many available channels are then selected uniformly without replacement.

Structural annotations use three repeat units joined by the declared single or double junction bond. SMARTS matches must intersect the central repeat. Polyether additionally requires an ether oxygen on a backbone route identified by the block-cut tree, allowing alternative paths through rings; pendant ethers can therefore satisfy ether but not polyether. The other motifs are defined by the rules in Table S3. Unsupported cases remain unassigned, and labels can overlap.

**Table S3.** Structural-control rules and illustrative repeat units. Labels may overlap.

| Label | Operational rule | Illustrative pSMILES |
|---|---|---|
| polyether | Backbone C-O-C ether oxygen satisfying the annotation exclusions | [*]CCO[*] |
| polycarbonate | O-C(=O)-O motif | [*]COC(=O)OC[*] |
| sulfone | Carbon-bound S(=O)2 | [*]CS(=O)(=O)C[*] |
| ester | Ester motif excluding carbonates and carbamates | [*]CCOC(=O)CC[*] |
| polysiloxane | Si-O-Si motif | [*]O[Si](C)(C)[*] |
| amide | Trivalent N linked to C(=O) | [*]CCNC(=O)CC[*] |
| imide | N linked to two carbonyl carbons | [*]CC(N1C(=O)CCC1=O)[*] |
| ether | Neutral C-O-C; carbonyl-adjacent exclusions | [*]CC(OCC)[*] |
| amine | Neutral amine; resonance/heteroatom exclusions | [*]CCN(CC)CC[*] |
| polyphosphazene | Alternating backbone P and N atoms with alternating single and double bonds | [*]P(OC)(OC)=N[*] |
| thioether | Neutral C-S-C; carbonyl-adjacent exclusions | [*]CCSCC[*] |

### S1.3 Training and optimization

Main-text Table 4 gives the model architecture, while Table S4 lists the training and software settings.

The SentencePiece vocabulary contains 265 inherited tokens and 388 tokens after adding controls and reserved symbols; we did not establish whether it was identical to the tokenizer distributed with polyBERT. Chemical sequences cannot contain unknown or control tokens. Prompts contain BOS, a generation-task token, optional electrolyte and structural-control tokens, a target marker, the chemical sequence, and EOS. One structural label may be requested for inclusion or exclusion. The numerical prefix and prompt tokens are excluded from prediction targets, and control and reserved tokens are excluded from chemical generation.

For structural conditioning, we randomized the order of the 22 possible inclusion and exclusion requests and shuffled the records without replacement during each pass. Each supervised chemical or EOS token receives weight $1/(M \times R_m \times N_{mr})$, where M is the number of modes present in the update, $R_m$ the number of requests present in mode m, and $N_{mr}$ the token count in that mode-request group. Modes without structural controls have one request. Weights use the full 1,024-example update before microbatch accumulation.

During training, we allowed one electrolyte example for every 138 accepted foundation examples and used each recognized electrolyte polymer group at most 44 times. Exact foundation overlaps counted toward this limit; unrecognized periodically equivalent representations did not.

By 43,000 optimizer updates, training had used 44,032,000 sampled examples and 3,668,604,753 polymer and end-of-sequence tokens for which the model was trained to make predictions. The balanced validation loss at this checkpoint was 0.255543. Training through 49,000 updates used 4,182,505,404 supervised tokens.

For each checkpoint, we calculated a validation score by averaging five cross-entropy losses with equal weight: unconditional, numerical, structural (averaged over requests), combined numerical-structural (averaged over requests), and electrolyte (averaged over polymer groups). This differs from the four-mode training loss and overall token cross-entropy.

**Table S4.** Optimization and software settings.

| Parameter | Setting |
|---|---|
| Dropout | 0.1 |
| Optimizer | AdamW; $\beta_1 = 0.9$, $\beta_2 = 0.95$, $\varepsilon = 10^{-8}$ |
| Batch size / microbatch size | 1,024 / 64 |
| Weight decay / gradient clipping | 0.1 for parameters with at least two dimensions; none for vectors or scalars / clipping at 1 |
| Peak / minimum learning rate | $3 \times 10^{-4}$ / $3 \times 10^{-5}$ |
| Warmup | 611 updates |

| Parameter | Setting |
|---|---|
| Validation interval | 500 optimizer updates |
| Plateau scheduler | Reduction factor 0.5; patience four validation evaluations; absolute improvement threshold $10^{-4}$; cooldown one evaluation |
| Stopping rule | At minimum learning rate, stop after twelve evaluations without improvement; reset the counter on improvement or learning-rate reduction |
| Training precision and hardware | bfloat16 on RTX 5070 Ti |
| Software | Python 3.11.9; PyTorch 2.11.0+cu128; NumPy 2.4.4; SentencePiece 0.1.99; RDKit 2026.3.3; NetworkX 3.6.1; pandas 3.0.2; PyArrow 24.0.0; protobuf 3.20.3 |

### S1.4 Property prediction

We used eight models based on pretrained TransPolymer [11] to predict properties of the generated polymer repeat units: chain and bulk electronic band gaps (Egc and Egb), electron affinity (Eea), crystallization tendency (Xc), glass-transition temperature (Tg), melting temperature (Tm), decomposition temperature (Td), and static dielectric constant (epsc). The electronic, Xc, and dielectric reference data originated from Kuenneth et al. [12]. Tg used the Kaggle compilation [13]; Tm and Td combined the data reported by Tang et al. [14], polyLearn [15], and OpenPoly [16], with additional Td records from PolyFKAN [17]. PoLyInfo [18] was an upstream source for records identified as originating from that database.

We trained five prediction models for each property using five-fold cross-validation. For Tg, Tm, and Td, different representations of the same polymer were kept in the same data subset. For Egc, Egb, Eea, and Xc, the data were divided by individual records, so different representations of the same polymer could appear in different subsets. For each model, the best-performing checkpoint was selected using its validation subset. When evaluating a generated polymer already present in the reference data, we used only models that had not been trained on that polymer group. For other generated polymers, we averaged predictions from all five models (Table S5).

The reported cross-validation errors describe model performance during development rather than independent validation of generated polymers. Some property labels used to train PolyGPT were also obtained from the predictors used for evaluation. Therefore, agreement with the requested properties does not necessarily indicate experimental accuracy. We did not establish calibrated uncertainty intervals for generated polymers. Furthermore, because experimental conditions are often incompletely reported for Tm and Td, the predictions cannot be assumed accurate under specific measurement conditions.

**Table S5.** Dataset sizes and performance of the property-prediction models. MAE and RMSE are in kelvin for Tg, Tm, and Td; electronvolts for Egc, Egb, and Eea; percentage points for Xc; and dimensionless units for the static dielectric constant. Xe has no assessment predictor.

| Property | Reference values | $R^2$ | MAE | RMSE |
|---|---|---|---|---|
| Tg | 7,247 | 0.905 | 21.720 | 34.832 |
| Tm | 3,913 | 0.761 | 35.420 | 52.644 |
| Td | 5,084 | 0.709 | 37.620 | 58.731 |
| Egc | 3,380 | 0.921 | 0.292 | 0.439 |
| Egb | 561 | 0.925 | 0.396 | 0.535 |
| Eea | 368 | 0.910 | 0.242 | 0.327 |
| Xc | 432 | 0.499 | 12.674 | 16.774 |
| Static dielectric (CV) | 300 | 0.738 | 0.382 | 0.566 |

### S1.5 Generation settings and scoring

Table 5 lists the generation conditions and the number of attempts for each evaluation. Unless varied in the sampling-sensitivity analysis, generation used temperature 1, top-k 0, top-p 1, batches of 64 with a smaller final batch where needed, and at most 398 new tokens. Inference used 32-bit floating-point arithmetic at the highest precision setting, with TensorFloat-32 disabled. All attempts for a given condition and seed were generated on one RTX 5070 Ti or RTX 2000 Ada GPU.

The principal and structural/domain evaluations used seeds 2026102811, 2026102812 and 2026102813; checkpoint comparisons used 2026103011, 2026103012 and 2026103013. A separate random stream was derived for each request. The principal low-$T_g$ target was 246.1134723721304 K, with all other numerical and structural conditions unspecified. Sensitivity-specific seeds and controls are given in Section S3.2.

The checkpoint comparison used 48 requests covering unconditional and electrolyte-only generation, eight Tg requests, sixteen other numerical requests and twenty-two single-label requests. Two Xe requests (29.009846% and 47.509566%) were generated but it remained unassessable since could not train any Xe predictor because of data unavailability. The other eight properties were assessed with the frozen predictors.

We classified an output as valid only if it ended with EOS and RDKit could parse it as a connected molecular graph with a nonempty interior, two terminal wildcard attachment sites, and compatible single or double bonds at the chain junction. Generation metrics and their denominators are defined in Table 6.

We calculated property errors from all valid outputs for which a prediction was available, including repeated generations. Numerical success requires every requested assessable property to lie within its acceptance band: ±50 K for thermal targets, ±0.5 eV for $E_{gc}/E_{gb}/E_{ea}$, ±10 percentage points for $X_c$, or ±1 unit for epsc. Any simultaneous structural request must also be satisfied by the same output. Outputs for which a required property could not be predicted were counted as unsuccessful. Structural-only tests use the compliance metrics defined in Table 6.

We calculated 95% Wilson confidence intervals for structural validity and useful-yield proportions. For the mean absolute error and signed bias, we resampled the scored outputs 1,000 times within each generation seed, including repeated structures. These intervals describe sampling variation within one trained model and exclude uncertainty from model training and property prediction.

For the molecular examples, we selected structures in the order generated, after checking their validity.

**S1.6 Polymer identity and novelty**

Repeat units were canonicalized with the Ramprasad open-pSMILES procedure and their attachment sites normalized. For periodic comparison, the shortest backbone path between the attachment neighbors was represented by its atom attributes, side branches and bond orders, including the closing junction bond. Both directions were reduced to their shortest repeating pattern and compared over all cyclic shifts. Structures with the same resulting representation were assigned to one conservative periodic group.

We applied the same grouping procedure to structures with missing annotations whenever their molecular graphs could be interpreted. If parsing, attachment-site assignment, or grouping failed, we marked the polymer identity as unknown. The backbone representation omits stereochemical tags and can merge backbone stereoisomers, although branch strings can retain local stereochemistry. It does not guarantee equivalence for rings with multiple backbone routes or complex repeat transformations. Canonical and periodic counts are therefore reported separately.

Training-catalog membership uses the reference defined in the main Methods. If an identifier was missing or a reference structure could not be normalized, we could not determine whether its canonical or conservative periodic identity was present in the training catalog. These outputs remain unknown and are excluded from novel counts. Periodic novelty is assessed separately from canonical property-compliant yields (Table 6).

## S2 Numerical and structural control

### S2.1 Numerical conditioning

Among the 5,808 electrolyte training groups, 1,728 had a retained $T_g$ label within 25 K of 246.11 K; none had a label within 50 K of 560.54 K. When we sampled electrolyte polymer groups equally and then sampled their repeat-unit variants equally, their mean retained Tg label was 291.3 K. These mostly predictor-derived labels indicate training support for low-$T_g$ requests.

**Table S6.** Single-property numerical control at 43,000 updates. Errors use valid, scored outputs; distinct useful yield uses all attempts and the tolerances in Section S1.5. Thermal errors are in kelvin, electronic errors in electronvolts, and $X_c$ errors in percentage points. All properties are predictor estimates. Only $T_g$ was evaluated with electrolyte conditioning.

| Condition | Target | Scored / attempted | MAE | RMSE | Bias | Distinct useful yield (%) |
|---|---|---|---|---|---|---|
| $E_{gc}$, lower | 2.88 | 3050/3072 | 0.31 | 0.39 | -0.17 | 80.9 |
| $E_{gc}$, upper | 3.97 | 3052/3072 | 0.38 | 0.47 | -0.31 | 70.0 |
| $E_{gb}$, lower | 2.49 | 3055/3072 | 0.39 | 0.50 | +0.06 | 70.7 |
| $E_{gb}$, upper | 3.64 | 3055/3072 | 0.61 | 0.74 | -0.43 | 45.1 |
| $E_{ea}$, lower | 1.87 | 3059/3072 | 0.36 | 0.45 | +0.27 | 72.9 |
| $E_{ea}$, upper | 2.65 | 3057/3072 | 0.26 | 0.32 | +0.14 | 87.9 |
| epsc, lower | 3.99 | 3062/3072 | 0.38 | 0.49 | +0.24 | 94.7 |
| epsc, upper | 4.90 | 3054/3072 | 0.40 | 0.51 | +0.02 | 94.2 |
| $X_c$, lower | 29.88 | 3057/3072 | 12.07 | 15.41 | +1.10 | 48.3 |
| $X_c$, upper | 52.84 | 3058/3072 | 16.77 | 19.46 | -4.34 | 29.6 |
| $T_d$, lower | 654.40 | 3045/3072 | 43.81 | 55.20 | -4.27 | 63.4 |
| $T_d$, upper | 771.75 | 3059/3072 | 46.03 | 59.42 | -30.45 | 63.4 |

| Condition | Target | Scored / attempted | MAE | RMSE | Bias | Distinct useful yield (%) |
|---|---|---|---|---|---|---|
| $T_g$, lower | 437.06 | 3051/3072 | 41.51 | 51.95 | +3.70 | 65.0 |
| $T_g$, upper | 560.54 | 3057/3072 | 31.30 | 39.82 | -3.62 | 80.0 |
| $T_m$, lower | 530.54 | 3051/3072 | 53.66 | 66.61 | -24.20 | 53.2 |
| $T_m$, upper | 653.91 | 3059/3072 | 49.12 | 63.50 | -32.33 | 59.4 |
| $T_g$ + electrolyte, lower | 437.06 | 3031/3072 | 67.61 | 72.56 | -64.42 | 15.0 |
| $T_g$ + electrolyte, central | 498.80 | 2964/3072 | 97.37 | 110.05 | -94.74 | 19.3 |
| $T_g$ + electrolyte, upper | 560.54 | 2943/3072 | 158.93 | 170.98 | -158.10 | 8.9 |
| $T_g$, central | 498.80 | 3057/3072 | 37.73 | 47.31 | -2.89 | 70.7 |
| Low $T_g$ | 246.11 | 3049/3072 | 26.53 | 33.63 | +4.10 | 86.1 |
| Low $T_g$ + electrolyte | 246.11 | 3061/3072 | 26.21 | 33.26 | +19.14 | 66.1 |

## S2.2 Structural and joint conditioning

**Table S7.** Valid and compliant yield for single structural requests at 43,000 updates. These requests omit electrolyte conditioning. Annotations can overlap.

| Structural annotation | Include yield (%) | Exclude yield (%) |
|---|---|---|
| polyether | 90.1 | 95.3 |
| polycarbonate | 99.1 | 98.9 |
| sulfone | 94.2 | 97.6 |
| ester | 90.5 | 95.0 |
| polysiloxane | 98.6 | 98.7 |
| amide | 91.8 | 94.1 |
| imide | 93.5 | 96.4 |
| ether | 90.5 | 95.7 |
| amine | 88.7 | 96.8 |
| polyphosphazene | 99.7 | 99.4 |
| thioether | 96.4 | 98.9 |

Three joint target sets were taken from individual training records near three standardized target centers. Two-property requests specify Tg and Egc; five-property requests add Egb, Eea and Xc. The target values came from property-prediction labels in the training data, not from new independent measurements. The corresponding joint-control results are reported in Table S9.

**Table S8.** Joint target sets A-C. Two-property requests use $T_g$ and $E_{gc}$; five-property requests use all five entries.

| Target set | $T_g$ (K) | $E_{gc}$ (eV) | $E_{gb}$ (eV) | $E_{ea}$ (eV) | $X_c$ (%) |
|---|---|---|---|---|---|
| A | 438.838 | 3.434 | 3.052 | 2.214 | 40.846 |
| B | 496.588 | 3.446 | 3.036 | 2.243 | 41.373 |
| C | 557.774 | 3.437 | 3.050 | 2.233 | 41.573 |

**Table S9.** Joint-control results at 43,000 updates, without the domain token. Success requires all requested numerical and structural conditions; distinct success removes canonical duplicates. All percentages use attempted generations. Tg-plus-ether requests specify Tg = 498.799 K.

| Condition | Valid (%) | Distinct (%) | All conditions (%) | Distinct all conditions (%) |
|---|---|---|---|---|
| 2 properties, set A | 99.5 | 99.5 | 49.9 | 49.9 |
| 5 properties, set A | 99.2 | 99.2 | 7.8 | 7.8 |
| 2 properties, set B | 99.2 | 99.2 | 48.8 | 48.8 |
| 5 properties, set B | 99.4 | 99.3 | 12.8 | 12.8 |
| 2 properties, set C | 99.5 | 99.5 | 58.1 | 58.1 |
| 5 properties, set C | 99.7 | 99.6 | 15.9 | 15.9 |
| $T_g$ + include ether | 99.3 | 99.3 | 64.9 | 64.9 |
| $T_g$ + exclude ether | 99.4 | 99.4 | 66.4 | 66.4 |

**Table S10.** Property predictions under joint conditions. Thermal quantities are in kelvin, electronic quantities in electronvolts, $X_c$ values in percent, and $X_c$ errors in percentage points. Bias is mean prediction minus target.

| Condition | Property | Requested | Mean prediction | MAE | Bias |
|---|---|---|---|---|---|
| 2 properties, set A | $E_{gc}$ | 3.434 | 3.256 | 0.346 | -0.178 |
| 2 properties, set A | $T_g$ | 438.838 | 447.513 | 41.237 | +8.675 |
| 5 properties, set A | $E_{gc}$ | 3.434 | 3.206 | 0.344 | -0.229 |
| 5 properties, set A | $E_{gb}$ | 3.052 | 2.835 | 0.444 | -0.217 |
| 5 properties, set A | $E_{ea}$ | 2.214 | 2.342 | 0.290 | +0.128 |
| 5 properties, set A | $X_c$ | 40.846 | 26.697 | 16.679 | -14.148 |
| 5 properties, set A | $T_g$ | 438.838 | 444.932 | 37.481 | +6.094 |
| 2 properties, set B | $E_{gc}$ | 3.446 | 3.171 | 0.363 | -0.275 |
| 2 properties, set B | $T_g$ | 496.588 | 495.696 | 38.866 | -0.892 |
| 5 properties, set B | $E_{gc}$ | 3.446 | 3.224 | 0.315 | -0.222 |
| 5 properties, set B | $E_{gb}$ | 3.036 | 2.813 | 0.481 | -0.223 |
| 5 properties, set B | $E_{ea}$ | 2.243 | 2.440 | 0.294 | +0.198 |
| 5 properties, set B | $X_c$ | 41.373 | 35.742 | 13.344 | -5.631 |
| 5 properties, set B | $T_g$ | 496.588 | 488.220 | 38.312 | -8.368 |
| 2 properties, set C | $E_{gc}$ | 3.437 | 3.089 | 0.380 | -0.348 |
| 2 properties, set C | $T_g$ | 557.774 | 552.750 | 28.657 | -5.024 |
| 5 properties, set C | $E_{gc}$ | 3.437 | 3.165 | 0.320 | -0.271 |
| 5 properties, set C | $E_{gb}$ | 3.050 | 2.688 | 0.522 | -0.362 |
| 5 properties, set C | $E_{ea}$ | 2.233 | 2.490 | 0.297 | +0.258 |
| 5 properties, set C | $X_c$ | 41.573 | 44.477 | 11.100 | +2.904 |
| 5 properties, set C | $T_g$ | 557.774 | 549.209 | 28.882 | -8.565 |
| $T_g$ + include ether | $T_g$ | 498.799 | 493.575 | 36.829 | -5.224 |
| $T_g$ + exclude ether | $T_g$ | 498.799 | 494.615 | 38.552 | -4.184 |

## S3 Effects of training duration and sampling

### S3.1 Training duration

To compare training checkpoints fairly, we generated structures with the same requests, random seeds, decoding settings, and sample sizes at each checkpoint (Table 5). Earlier checkpoints gave higher novel distinct useful yields for some requests (Figure 7 in the main manuscript and Table S11). Generation quality, $T_g$ error, and numerical and structural compliance across checkpoints are reported in Figure S1 and Tables S12, S13, S14, S15, and S16.

**Table S11.** Control across training checkpoints. Inclusion and exclusion average eleven requests each. Numerical yield averages lower and upper targets within each of eight properties, then weights the properties equally. These requests exclude electrolyte conditioning. Novel distinct useful yield is shown for $T_g$ = 246.11 K, with and without electrolyte conditioning, using a 50 K tolerance and all attempts as denominator.

| Updates (k) | Include (%) | Exclude (%) | Numerical (%) | Novel distinct useful $T_g$ (%) | Novel distinct useful $T_g$ + electrolyte (%) |
|---|---|---|---|---|---|
| 10 | 87.9 | 93.3 | 63.7 | 54.2 | 53.6 |
| 15 | 90.5 | 94.9 | 65.1 | 59.5 | 30.4 |
| 20 | 91.2 | 95.6 | 66.3 | 51.5 | 41.0 |
| 25 | 91.3 | 95.9 | 66.1 | 53.6 | 33.4 |
| 30 | 93.0 | 96.6 | 66.2 | 63.8 | 29.2 |
| 35 | 93.5 | 97.0 | 67.1 | 58.8 | 27.9 |
| 40 | 93.6 | 97.0 | 67.8 | 56.6 | 26.4 |
| 43 | 93.9 | 97.0 | 67.4 | 61.6 | 28.3 |
| 45 | 94.0 | 97.2 | 67.6 | 58.9 | 27.4 |

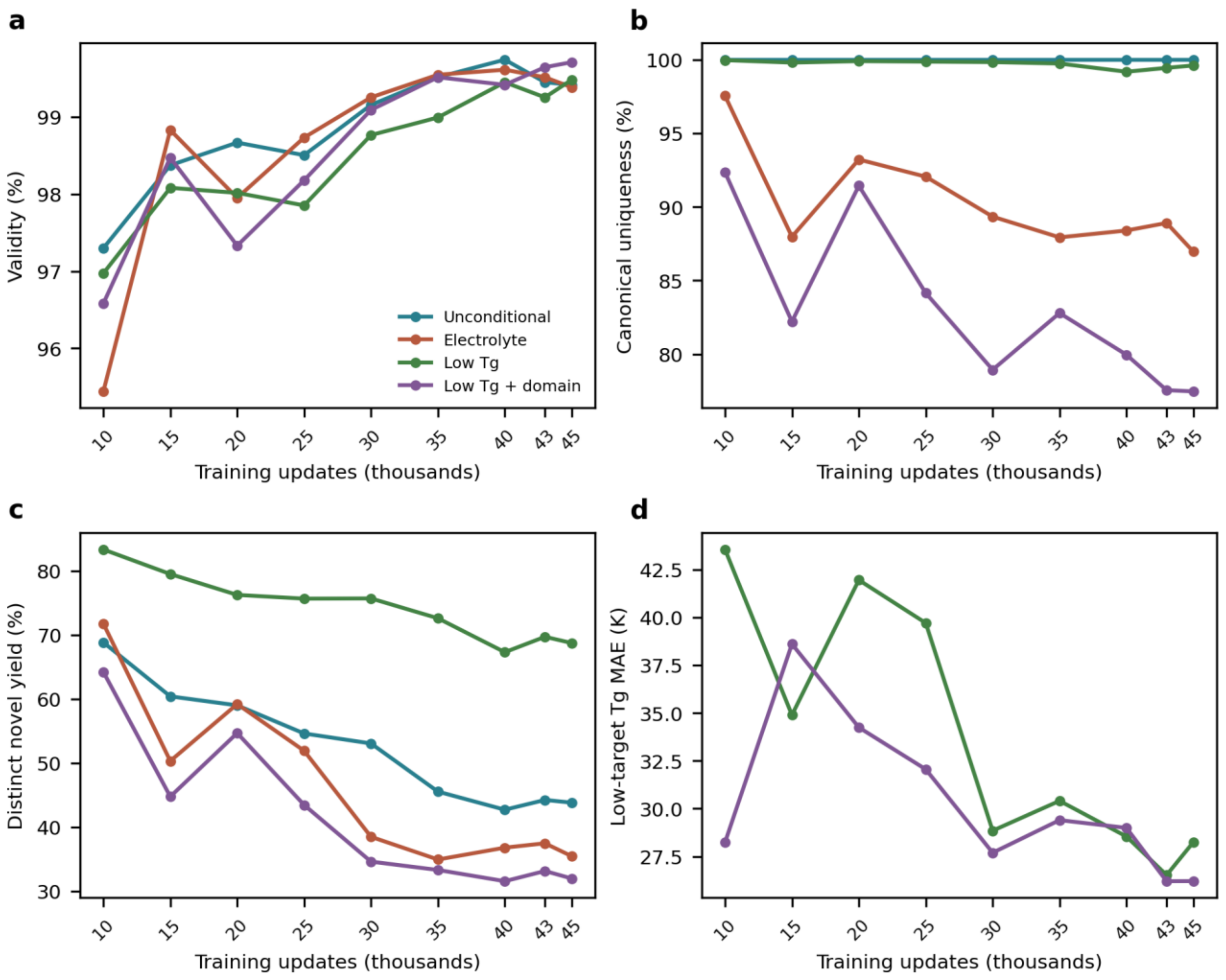

**Figure S1.** Training-duration dependence of (a) validity, (b) canonical uniqueness, (c) novel distinct yield, and (d) Tg MAE at 246.11 K. Novel distinct yield is the number of distinct canonical structures absent from the training catalog divided by all attempts. The checkpoints belong to one training run.

**Table S12.** Generation quality across training checkpoints. Uniqueness uses valid outputs; distinct and novel yields use all attempts. Tg MAE uses valid, scored outputs.

| Updates ($\times 10^3$) | Condition | Valid (%) | Unique / valid (%) | Distinct (%) | Novel distinct (%) | $T_g$ MAE (K) |
|---|---|---|---|---|---|---|
| 10 | Unconditional | 97.3 | 100.0 | 97.3 | 68.8 | — |
| | Electrolyte | 95.4 | 97.5 | 93.1 | 71.8 | — |
| | Low $T_g$ | 97.0 | 100.0 | 96.9 | 83.4 | 43.54 |

| Updates ($\times 10^3$) | Condition | Valid (%) | Unique / valid (%) | Distinct (%) | Novel distinct (%) | $T_g$ MAE (K) |
|---|---|---|---|---|---|---|
| | Low $T_g$ + electrolyte | 96.6 | 92.3 | 89.2 | 64.2 | 28.24 |
| 15 | Unconditional | 98.4 | 100.0 | 98.4 | 60.4 | — |
| | Electrolyte | 98.8 | 88.0 | 86.9 | 50.3 | — |
| | Low $T_g$ | 98.1 | 99.8 | 97.9 | 79.5 | 34.89 |
| | Low $T_g$ + electrolyte | 98.5 | 82.2 | 81.0 | 44.8 | 38.60 |
| 20 | Unconditional | 98.7 | 100.0 | 98.7 | 59.0 | — |
| | Electrolyte | 97.9 | 93.2 | 91.3 | 59.2 | — |
| | Low $T_g$ | 98.0 | 99.9 | 97.9 | 76.3 | 41.95 |
| | Low $T_g$ + electrolyte | 97.3 | 91.4 | 89.0 | 54.7 | 34.24 |
| 25 | Unconditional | 98.5 | 100.0 | 98.5 | 54.6 | — |
| | Electrolyte | 98.7 | 92.1 | 90.9 | 51.9 | — |
| | Low $T_g$ | 97.9 | 99.9 | 97.7 | 75.7 | 39.70 |
| | Low $T_g$ + electrolyte | 98.2 | 84.2 | 82.6 | 43.5 | 32.06 |
| 30 | Unconditional | 99.2 | 100.0 | 99.2 | 53.1 | — |
| | Electrolyte | 99.3 | 89.3 | 88.7 | 38.5 | — |
| | Low $T_g$ | 98.8 | 99.8 | 98.6 | 75.7 | 28.85 |
| | Low $T_g$ + electrolyte | 99.1 | 78.9 | 78.2 | 34.6 | 27.70 |
| 35 | Unconditional | 99.5 | 100.0 | 99.5 | 45.5 | — |
| | Electrolyte | 99.5 | 87.9 | 87.5 | 34.9 | — |
| | Low $T_g$ | 99.0 | 99.7 | 98.7 | 72.6 | 30.41 |
| | Low $T_g$ + electrolyte | 99.5 | 82.8 | 82.4 | 33.3 | 29.40 |
| 40 | Unconditional | 99.7 | 100.0 | 99.7 | 42.7 | — |
| | Electrolyte | 99.6 | 88.4 | 88.1 | 36.8 | — |
| | Low $T_g$ | 99.4 | 99.2 | 98.6 | 67.3 | 28.54 |
| | Low $T_g$ + electrolyte | 99.4 | 80.0 | 79.5 | 31.5 | 28.99 |
| 43 | Unconditional | 99.4 | 100.0 | 99.4 | 44.2 | — |
| | Electrolyte | 99.5 | 88.9 | 88.5 | 37.5 | — |
| | Low $T_g$ | 99.3 | 99.4 | 98.7 | 69.7 | 26.53 |
| | Low $T_g$ + electrolyte | 99.6 | 77.6 | 77.3 | 33.1 | 26.21 |
| 45 | Unconditional | 99.4 | 100.0 | 99.4 | 43.8 | — |

| Updates (×10³) | Condition | Valid (%) | Unique / valid (%) | Distinct (%) | Novel distinct (%) | $T_g$ MAE (K) |
|---|---|---|---|---|---|---|
| | Electrolyte | 99.4 | 87.0 | 86.4 | 35.4 | — |
| | Low $T_g$ | 99.5 | 99.6 | 99.1 | 68.8 | 28.25 |
| | Low $T_g$ + electrolyte | 99.7 | 77.5 | 77.2 | 32.0 | 26.21 |

**Table S13.** Tg-only MAE (K) across checkpoints. Errors use valid, scored outputs.

| Updates (k) | 246.11 K | 437.06 K | 498.80 K | 560.54 K |
|---|---|---|---|---|
| 10 | 43.54 | 45.28 | 40.01 | 35.81 |
| 15 | 34.89 | 44.20 | 41.12 | 32.39 |
| 20 | 41.95 | 41.94 | 38.56 | 31.44 |
| 25 | 39.70 | 43.27 | 40.79 | 33.09 |
| 30 | 28.85 | 42.53 | 38.50 | 32.16 |
| 35 | 30.41 | 42.29 | 38.59 | 30.93 |
| 40 | 28.54 | 41.06 | 37.54 | 30.54 |
| 43 | 26.53 | 41.51 | 37.73 | 31.30 |
| 45 | 28.25 | 41.29 | 38.45 | 31.90 |

**Table S14.** Tg MAE (K) with electrolyte conditioning, using the targets in Table S13.

| Updates (k) | 246.11 K | 437.06 K | 498.80 K | 560.54 K |
|---|---|---|---|---|
| 10 | 28.24 | 102.18 | 150.43 | 212.25 |
| 15 | 38.60 | 92.15 | 137.53 | 189.94 |
| 20 | 34.24 | 77.09 | 123.68 | 175.83 |
| 25 | 32.06 | 85.10 | 130.17 | 189.07 |
| 30 | 27.70 | 80.31 | 131.46 | 192.32 |
| 35 | 29.40 | 70.56 | 119.09 | 177.31 |
| 40 | 28.99 | 66.90 | 101.84 | 162.73 |
| 43 | 26.21 | 67.61 | 97.37 | 158.93 |
| 45 | 26.21 | 67.57 | 96.15 | 159.51 |

**Table S15.** Property compliance across checkpoints without electrolyte conditioning. Each percentage equally averages lower and upper target yields. Invalid or unscorable attempts fail. Tolerances are defined in Section S1.5; $X_e$ is unassessed.

| Updates (k) | $E_{gc}$ | $E_{gb}$ | $E_{ea}$ | epsc | $X_c$ | $T_d$ | $T_g$ | $T_m$ |
|---|---|---|---|---|---|---|---|---|
| 10 | 72.7 | 53.7 | 76.5 | 92.9 | 38.8 | 57.6 | 66.4 | 51.2 |
| 15 | 72.6 | 57.8 | 77.9 | 92.8 | 38.6 | 58.0 | 69.9 | 53.2 |
| 20 | 72.2 | 55.9 | 81.0 | 93.9 | 39.0 | 61.6 | 71.5 | 55.6 |
| 25 | 73.7 | 56.3 | 79.1 | 94.3 | 38.4 | 62.5 | 70.8 | 53.4 |
| 30 | 72.7 | 56.5 | 79.9 | 94.0 | 39.7 | 62.2 | 71.4 | 53.3 |
| 35 | 74.6 | 57.8 | 80.0 | 94.4 | 39.9 | 62.4 | 72.4 | 55.4 |
| 40 | 76.4 | 56.9 | 80.0 | 94.6 | 39.3 | 64.6 | 73.6 | 56.6 |
| 43 | 75.4 | 57.9 | 80.4 | 94.5 | 39.0 | 63.4 | 72.5 | 56.3 |
| 45 | 76.1 | 57.8 | 80.4 | 94.6 | 39.9 | 63.8 | 72.7 | 55.6 |

**Table S16.** Structural compliance across checkpoints. Each entry is the percentage of all attempts satisfying the indicated request without electrolyte conditioning. Updates are in thousands.

| Request | 10 | 15 | 20 | 25 | 30 | 35 | 40 | 43 | 45 |
|---|---|---|---|---|---|---|---|---|---|
| Include polyether | 82.3 | 87.7 | 85.9 | 86.1 | 88.5 | 88.8 | 90.2 | 90.1 | 89.8 |
| Include polycarbonate | 96.1 | 97.8 | 97.5 | 96.8 | 98.5 | 98.7 | 98.9 | 99.1 | 98.9 |
| Include sulfone | 83.6 | 85.5 | 88.9 | 90.3 | 91.2 | 92.4 | 93.0 | 94.2 | 93.2 |
| Include ester | 80.3 | 80.7 | 83.9 | 85.0 | 89.3 | 88.6 | 90.2 | 90.5 | 90.4 |
| Include polysiloxane | 94.4 | 96.9 | 96.8 | 98.1 | 97.4 | 98.2 | 98.6 | 98.6 | 98.4 |
| Include amide | 88.2 | 91.5 | 93.2 | 90.7 | 92.1 | 94.4 | 92.5 | 91.8 | 93.7 |
| Include imide | 88.4 | 90.7 | 93.7 | 91.8 | 94.6 | 95.1 | 92.4 | 93.5 | 93.9 |
| Include ether | 85.0 | 89.5 | 86.8 | 86.0 | 88.8 | 88.3 | 89.7 | 90.5 | 89.6 |
| Include amine | 81.4 | 83.9 | 84.2 | 86.0 | 87.9 | 89.2 | 88.2 | 88.7 | 89.0 |
| Include polyphosphazene | 98.6 | 99.3 | 99.4 | 99.2 | 99.6 | 99.8 | 99.9 | 99.7 | 99.9 |

| Request | 10 | 15 | 20 | 25 | 30 | 35 | 40 | 43 | 45 |
|---|---|---|---|---|---|---|---|---|---|
| Include thioether | 89.1 | 92.2 | 92.8 | 94.3 | 95.1 | 95.4 | 95.5 | 96.4 | 96.8 |
| Exclude polyether | 90.5 | 93.0 | 93.8 | 95.3 | 95.8 | 96.1 | 96.1 | 95.3 | 96.6 |
| Exclude polycarbonate | 97.6 | 97.9 | 98.4 | 98.1 | 99.0 | 99.0 | 98.9 | 98.9 | 98.9 |
| Exclude sulfone | 94.5 | 96.5 | 97.2 | 96.8 | 98.0 | 98.3 | 98.0 | 97.6 | 98.3 |
| Exclude ester | 89.6 | 91.2 | 93.6 | 94.1 | 94.2 | 94.7 | 94.9 | 95.0 | 95.4 |
| Exclude polysiloxane | 97.2 | 98.1 | 97.4 | 97.9 | 98.3 | 98.6 | 98.8 | 98.7 | 99.0 |
| Exclude amide | 88.9 | 91.5 | 91.5 | 91.6 | 93.5 | 94.3 | 94.9 | 94.1 | 94.9 |
| Exclude imide | 92.4 | 94.0 | 95.8 | 94.6 | 96.2 | 97.0 | 96.7 | 96.4 | 96.4 |
| Exclude ether | 91.6 | 92.7 | 94.6 | 95.2 | 95.2 | 95.9 | 96.2 | 95.7 | 95.7 |
| Exclude amine | 91.3 | 94.6 | 94.4 | 95.0 | 95.6 | 96.6 | 96.2 | 96.8 | 96.7 |
| Exclude polyphosphazene | 97.6 | 98.2 | 98.7 | 98.7 | 98.8 | 99.1 | 99.1 | 99.4 | 99.1 |
| Exclude thioether | 94.6 | 95.6 | 96.5 | 97.2 | 97.7 | 97.6 | 97.9 | 98.9 | 98.3 |

### S3.2 Sampling sensitivity at 43,000 updates

The sampling-sensitivity grid and sample sizes are given in Table 5. A top-k value of 0 allowed the model to sample from the full token vocabulary, while top-p was held at 1. The low-$T_g$-plus-electrolyte target was 246.113472 K, with other numerical and structural conditions unspecified.

Sensitivity sampling used seeds 2026102821, 2026102822 and 2026102823. For each condition and seed, the derived random seed and physical GPU were held fixed across sampling settings. Other generation and assessment procedures followed Section S1.5. We kept the sampling-sensitivity results separate from the principal libraries and the nine-checkpoint comparison. The principal decoding settings were not changed using these results.

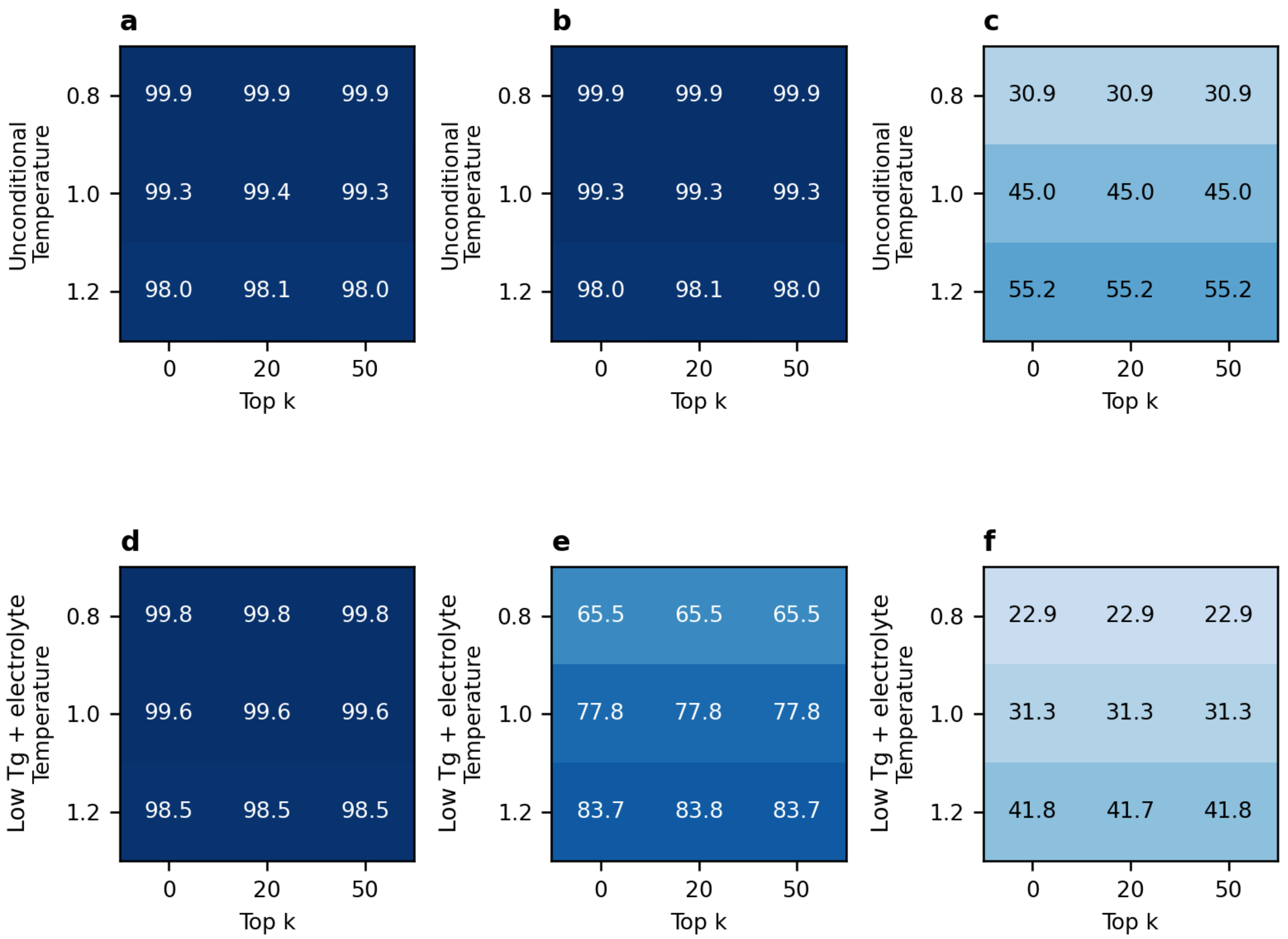

**Figure S2.** Sampling sensitivity for (a-c) unconditional and (d-f) low-Tg-plus-electrolyte generation. Columns show validity, distinct yield and novel distinct yield. Each cell pools the three seeds at top-p 1; all percentages use attempted generations.

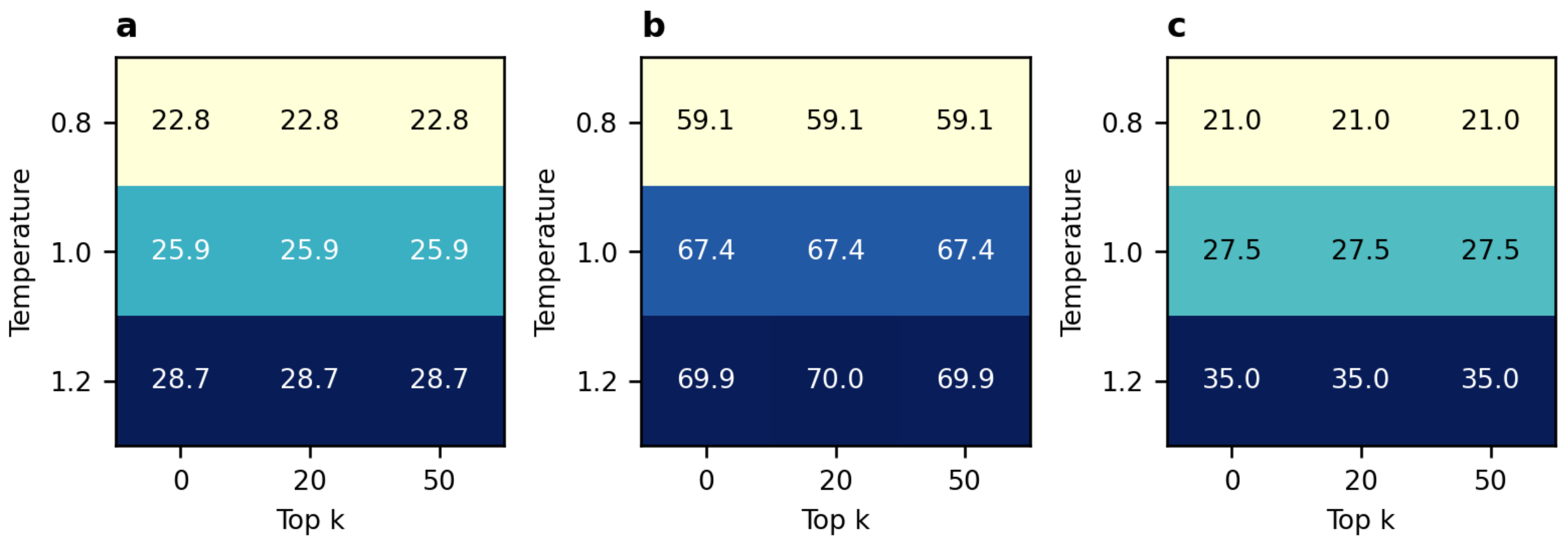

**Figure S3.** Sampling sensitivity of low-Tg-plus-electrolyte generation: (a) Tg MAE, (b) distinct useful yield, and (c) novel distinct useful yield. Useful outputs in panels (b,c) have a predicted $T_g$ within ±50 K of the 246.113472 K target.

## S4 Additional characterization of generated libraries

### S4.1 Library growth and replicate variation

The principal evaluation yielded 119,336 valid outputs and 80,696 distinct canonical structures, all with Tg predictions. Training-set membership was resolved for all valid outputs. For a structure that matched a known reference polymer group, property prediction used only models trained without that group, as described in Section S1.4. Generation identifiers and predictions for the first valid examples shown in Figure 9.

**Table S17.** Generation quality for three sampling replicates per condition. Pooled counts remove duplicates across replicates and therefore differ from sums of these distinct counts.

| Condition | Replicate | Valid | Canonical distinct | Periodic distinct | Unique (%) | Novel distinct |
|---|---|---|---|---|---|---|
| Unconditional | 1 | 9,942 | 9,941 | 9,941 | 99.99 | 4,387 |
| | 2 | 9,948 | 9,948 | 9,948 | 100.00 | 4,478 |
| | 3 | 9,947 | 9,947 | 9,947 | 100.00 | 4,324 |
| Electrolyte | 1 | 9,964 | 7,250 | 6,473 | 72.76 | 3,383 |

| Condition | Replicate | Valid | Canonical distinct | Periodic distinct | Unique (%) | Novel distinct |
|---|---|---|---|---|---|---|
| | 2 | 9,953 | 7,188 | 6,442 | 72.22 | 3,430 |
| | 3 | 9,959 | 7,197 | 6,419 | 72.27 | 3,341 |
| Low $T_g$ | 1 | 9,924 | 9,767 | 9,729 | 98.42 | 6,921 |
| | 2 | 9,931 | 9,781 | 9,738 | 98.49 | 6,870 |
| | 3 | 9,919 | 9,775 | 9,747 | 98.55 | 6,990 |
| Low $T_g$ + electrolyte | 1 | 9,953 | 5,764 | 4,919 | 57.91 | 2,901 |
| | 2 | 9,954 | 5,805 | 4,966 | 58.32 | 2,899 |
| | 3 | 9,942 | 5,750 | 4,916 | 57.84 | 2,844 |

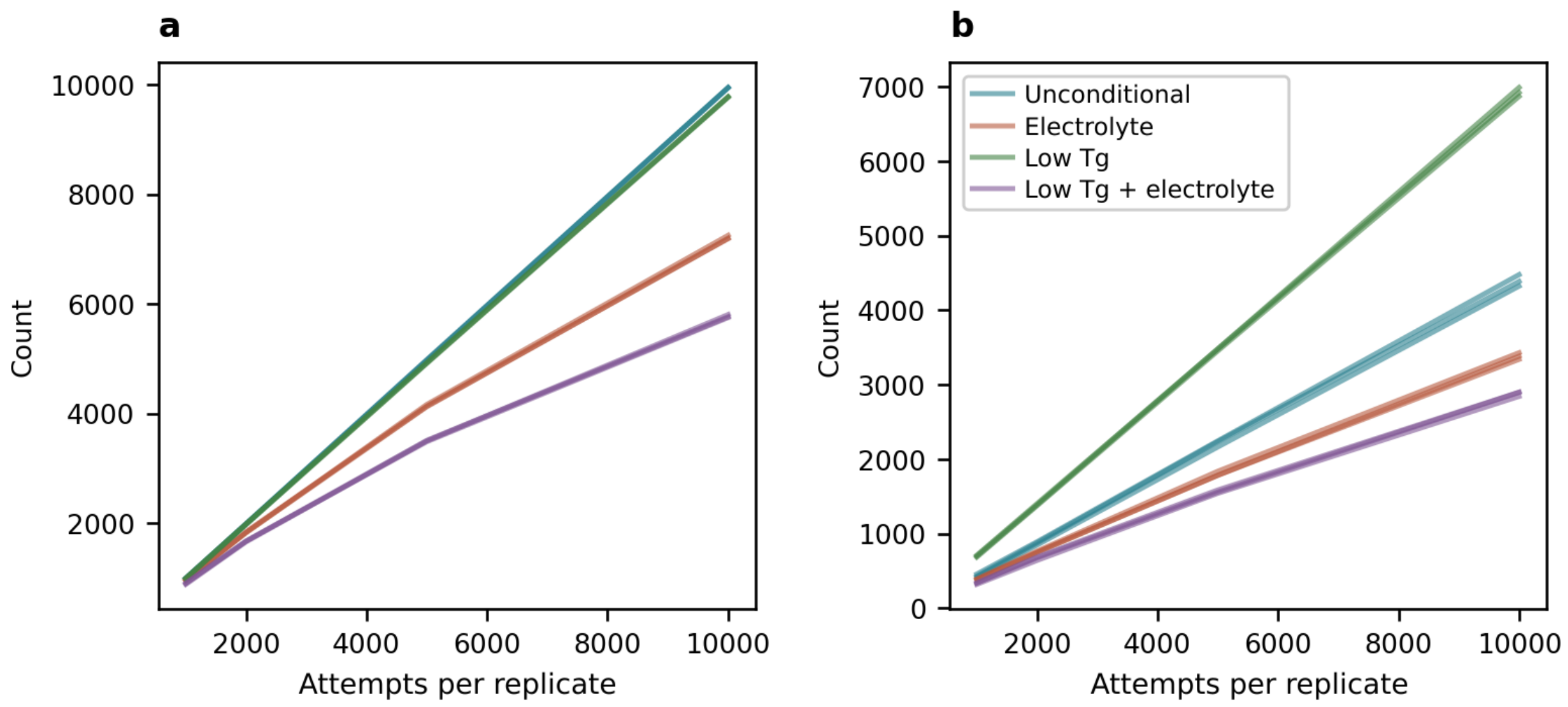

**Figure S4.** Accumulation of (a) distinct valid and (b) novel distinct structures across nested generation prefixes. The three lines per condition correspond to the three generation replicates.

### S4.2 $T_g$ tolerance and prediction error

For the principal libraries, we counted how many generated structures had predicted Tg values within 25 K or 50 K of the requested 246.11 K (Figure 8 and Table 3 in the main manuscript).

For the fraction of all attempts yielding valid, scored outputs within 25 K of the target, the 95% Wilson confidence intervals were 52.89-54.02% for Tg-only generation and 56.10-57.22% for Tg plus electrolyte conditioning. At 50 K, the corresponding intervals were 85.49-86.28% and 86.86-87.62%, respectively. These intervals apply before canonical deduplication.

**Table S18.** Tg error by replicate in the principal libraries, at a target of 246.11 K. Percentile intervals use 1,000 bootstrap resamples of valid, scored draws, including repeats.

| Condition | Replicate | MAE (K) | MAE 95% interval | Bias (K) | Bias 95% interval |
|---|---|---|---|---|---|
| Low $T_g$ | 1 | 27.01 | 26.62-27.45 | +5.11 | 4.47-5.75 |
| | 2 | 26.81 | 26.41-27.18 | +5.06 | 4.39-5.68 |
| | 3 | 27.10 | 26.69-27.52 | +5.12 | 4.44-5.80 |
| Low $T_g$ + electrolyte | 1 | 26.12 | 25.72-26.52 | +19.07 | 18.54-19.59 |
| | 2 | 26.14 | 25.75-26.52 | +19.49 | 19.00-20.00 |
| | 3 | 25.94 | 25.56-26.33 | +19.36 | 18.84-19.87 |

### S4.3 Periodic representations

**Table S19.** Counts under canonical and conservative periodic identities. Novel counts require absence from the training set. No Tg tolerance is applied to these counts.

| Condition | Canonical distinct | Periodic distinct | Canonical novel | Novel periodic groups |
|---|---|---|---|---|
| Unconditional | 29,835 | 29,832 | 13,189 | 12,980 |
| Electrolyte | 15,458 | 12,881 | 8,745 | 7,934 |
| Low $T_g$ | 28,588 | 28,326 | 20,621 | 20,518 |
| Low $T_g$ + electrolyte | 11,664 | 9,580 | 6,964 | 6,159 |

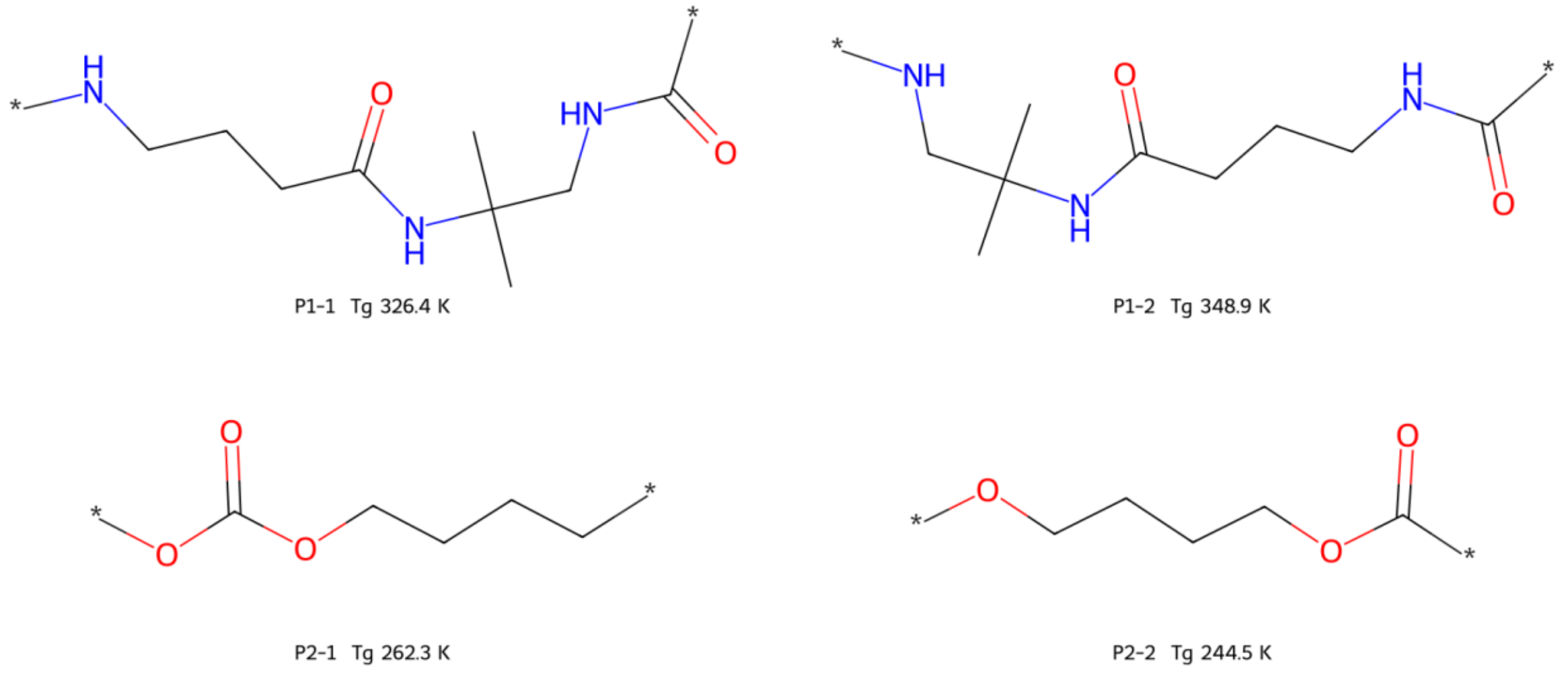

**Figure S5.** Two canonical representations sharing a conservative periodic identity for electrolyte-only (P1) and low-Tg-plus-electrolyte (P2) generation. The first pair encountered in each condition was used. Displayed Tg values are predictor estimates.

## S5 Sequence likelihood and training curves

After choosing the 43,000-update checkpoint, we evaluated how accurately it predicted the tokens in 5,074 held-out polymer sequences and their associated prompts. Overall token cross-entropy was 0.207279, the balanced loss was 0.246207, and electrolyte cross-entropy averaged equally across groups was 0.422500 (Table S20).

The test used foundation examples from each conditioning mode and all 58 electrolyte test variants. For the two structural conditioning modes, we presented the same source sequences with different structural requests.

Test examples came from the designated source-data partition; electrolyte groups were also disjoint across partitions. These partitions do not exclude the possibility that another representation of the same polymer appeared in the tokenizer-training data or a property-predictor dataset. Sequence likelihood assesses the sequence model rather than the physical properties of generated polymers.

**Table S20.** Test-set sequence likelihood. Token cross-entropy weights tokens equally within each conditioning family; the balanced objective follows the validation weighting in Section S1.3.

| Condition family | Examples | Supervised tokens | Token cross-entropy |
|---|---|---|---|
| Unconditional | 2,163 | 186,098 | 0.230455 |
| Numerical | 2,149 | 181,556 | 0.186585 |
| Structural only | 352 | 29,484 | 0.216431 |
| Numerical + structural | 352 | 29,484 | 0.167829 |
| Electrolyte | 58 | 1,604 | 0.417669 |
| Overall | 5,074 | 428,226 | 0.207279 |

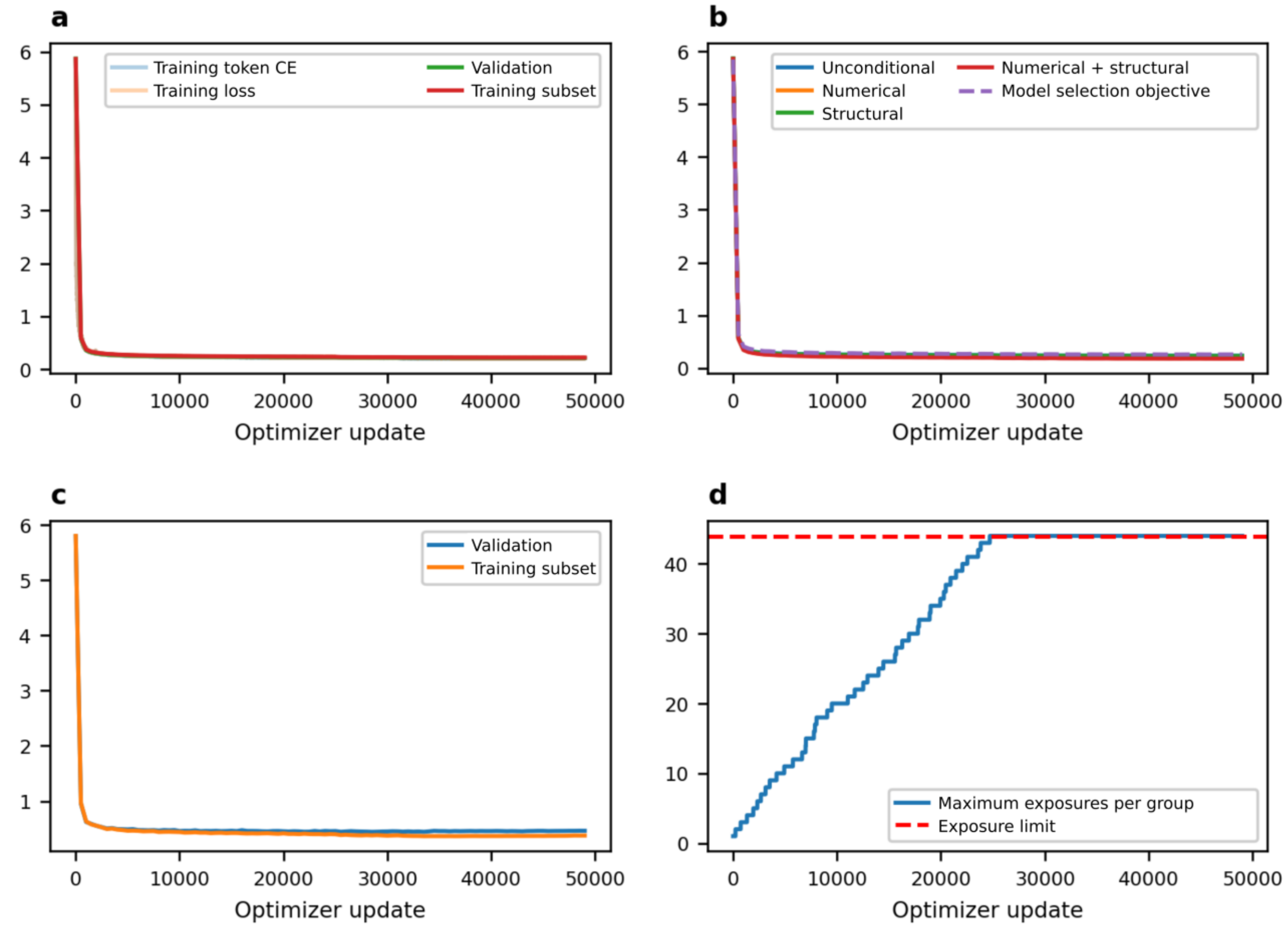

**Figure S6.** Training and validation through 49,000 updates: (a) token cross-entropy and balanced training loss, (b) validation by conditioning mode, (c) electrolyte cross-entropy averaged across groups, and (d) electrolyte-group exposure. The 43,000-update model minimizes balanced validation loss. The training subset denotes examples used to monitor training loss.

## S6 Chemical space and composition

The four principal libraries were analyzed after pooling the three generation seeds and removing canonical duplicates within each condition. We counted each distinct valid structure once in each pooled library. We also removed duplicates separately within each seed, so adding the per-seed distinct counts does not give the pooled total. Binary, chirality-aware, radius-two Morgan fingerprints with 2,048 bits represented finite repeat units with wildcard attachment points. The reference comprised structures encountered in training updates 1-43,000, excluding validation, test and later training records. Deduplication by RDKit canonical isomeric SMILES yielded 34,263,511 finite repeat-unit graphs, weighted equally rather than by exposure. This encountered reference is distinct from the full eligible training catalog used for novelty in Section S1.6.

Mean pairwise similarity follows the averaging approach of Yang et al. [6]. For each pooled condition, three million uniformly sampled training-generated pairs were evaluated using three pair-sampling seeds. Each generation-seed estimate used sampled pairs whose generated structure belonged to that seed's deduplicated library; the pooled result was calculated directly from all sampled pairs rather than by averaging the three seed means. The reference was not restricted to a property bin or electrolyte subset, and training matches were included. Secondary estimates excluded exact finite canonical matches in the encountered catalog. The reported Monte Carlo intervals describe uncertainty from sampling fingerprint pairs, not differences between separately trained models. Similarities were independently recomputed from all recorded integer intersection and union counts, with additional RDKit checks on selected pairs. This follows the averaging approach, without exactly reproducing the endpoint encoding, chirality defaults or reference collection of Yang et al. Means and intervals appear in Tables S22 and S25.

For the nearest-neighbour analysis in Figure S7, every generated query was compared with the complete encountered reference without approximate neighbour indexing. Computed fingerprint intersections were checked against CPU calculations, and every maximum-similarity score was independently verified using bitwise intersections and unions. For 16 queries, we also searched the complete reference using fingerprint bit counts and checked selected results with RDKit.

Internal diversity is one minus the mean Tanimoto similarity over 10,000 pairs sampled with replacement, including possible self-pairs. Sampling used sorted canonical representations and NumPy random seed 20261027. Functional-group prevalence excluded unknown annotations from its denominator; descriptors, annotation coverage and per-seed results are reported in Tables S21, S23 and S24.

The PCA basis was fitted in 64-bit arithmetic to a fixed 1,024-record subset of the eligible training data, without generated structures; the first two components explained 8.6% of the fitting-set variance. Generated and encountered-reference structures were projected without refitting. Gaussian densities used a shared bandwidth matrix estimated from pooled generated coordinates by Scott's rule. Generated densities were evaluated on a 512 × 512 grid; the training-reference grid used the same spacing and covered the complete reference and kernel tails. Each density was normalized separately, and contours enclosed 50%, 80% and 95% of its probability mass. The contours indicate the fraction of the estimated density enclosed, not uncertainty in that density.

**Table S21.** Pooled chemical-space descriptors. Heavy-atom counts exclude wildcard attachment atoms. Internal diversity is one minus mean Tanimoto similarity over 10,000 sampled pairs. Nearest-reference similarity is the maximum over all 34,263,511 distinct structures encountered through update 43,000, not the full eligible training catalog.

| Condition | Distinct valid | Median heavy atoms | Internal diversity | Median reference similarity | Unknown annotations |
|---|---|---|---|---|---|
| Unconditional | 29,835 | 42 | 0.8386 | 0.9038 | 1,365 |
| Electrolyte | 15,458 | 14 | 0.7636 | 0.7500 | 25 |
| Low Tg | 28,588 | 26 | 0.8150 | 0.8958 | 449 |
| Low Tg + domain | 11,664 | 14 | 0.7485 | 0.7561 | 36 |

**Table S22.** Mean pairwise Tanimoto similarity for pooled distinct valid candidates. The 95% interval describes Monte Carlo sampling error. The nonmatching mean excludes exact finite canonical matches in the encountered reference, not all eligible training records.

| Condition | Mean pairwise similarity | Monte Carlo 95% interval | Nonmatching mean |
|---|---|---|---|
| Unconditional | 0.159965 | 0.159891 to 0.160039 | 0.158960 |
| Electrolyte | 0.066345 | 0.066310 to 0.066381 | 0.066559 |
| Low Tg | 0.091488 | 0.091432 to 0.091544 | 0.091082 |
| Low Tg + domain | 0.065661 | 0.065625 to 0.065697 | 0.065637 |

Electrolyte conditioning reduces median repeat-unit size and within-library fingerprint diversity. Low-Tg generation without domain conditioning is intermediate in both. A library may contain more similar structures within itself while being less similar, on average, to the polymers encountered during training. The two-component projection does not establish full chemical-space coverage.

**Table S23.** Functional-group prevalence among assessable distinct canonical structures, in percent. Columns use the pooled libraries in Table S21. Known-annotation denominators are 28,470, 15,433, 28,139 and 11,628, respectively. Unknown annotations are excluded, not classified as absent. Labels overlap and do not sum to 100%.

| Annotation | Unconditional | Electrolyte | Low Tg | Low Tg + domain |
|---|---|---|---|---|
| polyether | 28.35 | 4.85 | 47.19 | 14.36 |
| polycarbonate | 0.40 | 56.41 | 5.65 | 74.56 |
| sulfone | 9.39 | 0.70 | 0.37 | 0.51 |
| ester | 10.47 | 7.06 | 46.01 | 11.40 |
| polysiloxane | 0.53 | 0.05 | 10.98 | 0.28 |
| amide | 61.55 | 59.10 | 18.94 | 30.70 |
| imide | 54.16 | 0.25 | 4.25 | 0.12 |
| ether | 35.54 | 15.17 | 59.17 | 27.64 |
| amine | 22.86 | 41.99 | 15.98 | 46.70 |
| polyphosphazene | 0.10 | 0.00 | 3.19 | 0.00 |
| thioether | 3.46 | 3.79 | 9.71 | 6.66 |

**Table S24.** Chemical-space descriptors for each generation seed. Each seed is deduplicated separately; values are not weighted by generation frequency. Nearest-reference similarity uses the same complete encountered reference as Table S21.

| Condition | Seed | Distinct valid | Median heavy atoms | Internal diversity | Median reference similarity | Unknown annotations |
|---|---|---|---|---|---|---|
| Unconditional | 2026102811 | 9,941 | 42 | 0.8398 | 0.9032 | 459 |
| | 2026102812 | 9,948 | 42 | 0.8377 | 0.9016 | 458 |
| | 2026102813 | 9,947 | 42 | 0.8396 | 0.9077 | 448 |
| Electrolyte | 2026102811 | 7,250 | 14 | 0.7594 | 0.7838 | 8 |
| | 2026102812 | 7,188 | 14 | 0.7620 | 0.7812 | 6 |
| | 2026102813 | 7,197 | 14 | 0.7588 | 0.7838 | 11 |
| Low Tg | 2026102811 | 9,767 | 26 | 0.8151 | 0.9000 | 144 |
| | 2026102812 | 9,781 | 26 | 0.8139 | 0.9118 | 157 |
| | 2026102813 | 9,775 | 26 | 0.8139 | 0.9024 | 150 |
| Low Tg + domain | 2026102811 | 5,764 | 14 | 0.7408 | 0.7941 | 12 |
| | 2026102812 | 5,805 | 14 | 0.7398 | 0.7949 | 16 |
| | 2026102813 | 5,750 | 14 | 0.7423 | 0.7949 | 16 |

**Table S25.** Mean pairwise Tanimoto similarity for each generation seed. The intervals describe Monte Carlo sampling error within a generated library.

| Condition | Generation seed | Mean pairwise similarity | Monte Carlo 95% interval |
|---|---|---|---|
| Unconditional | 2026102811 | 0.159898 | 0.159770 to 0.160025 |
| Unconditional | 2026102812 | 0.160304 | 0.160176 to 0.160432 |
| Unconditional | 2026102813 | 0.159689 | 0.159561 to 0.159817 |
| Electrolyte | 2026102811 | 0.066180 | 0.066128 to 0.066231 |
| Electrolyte | 2026102812 | 0.066083 | 0.066031 to 0.066135 |
| Electrolyte | 2026102813 | 0.066260 | 0.066209 to 0.066312 |
| Low Tg | 2026102811 | 0.091497 | 0.091402 to 0.091592 |
| Low Tg | 2026102812 | 0.091094 | 0.090999 to 0.091189 |
| Low Tg | 2026102813 | 0.091080 | 0.090985 to 0.091175 |
| Low Tg + domain | 2026102811 | 0.065683 | 0.065632 to 0.065735 |
| Low Tg + domain | 2026102812 | 0.065633 | 0.065582 to 0.065683 |
| Low Tg + domain | 2026102813 | 0.065659 | 0.065608 to 0.065710 |

Canonical representation dependence, fingerprint collisions and the limited variance captured by the two-component projection remain limitations. Figure S7 uses fixed 0.05-wide bins for similarities below one and displays exact-one fractions separately. Both use all distinct candidates in each condition as the denominator; the below-one subset is not renormalized. Median markers include exact-one values. Histogram and exact-one bars have separately labelled percentage scales. The fractions at one are 34.71%, 28.71%, 40.12% and 26.45% for unconditional, electrolyte-only, low-Tg and combined generation, respectively. These percentages count identical fingerprints; they do not establish identical canonical structures or memorization by the model.

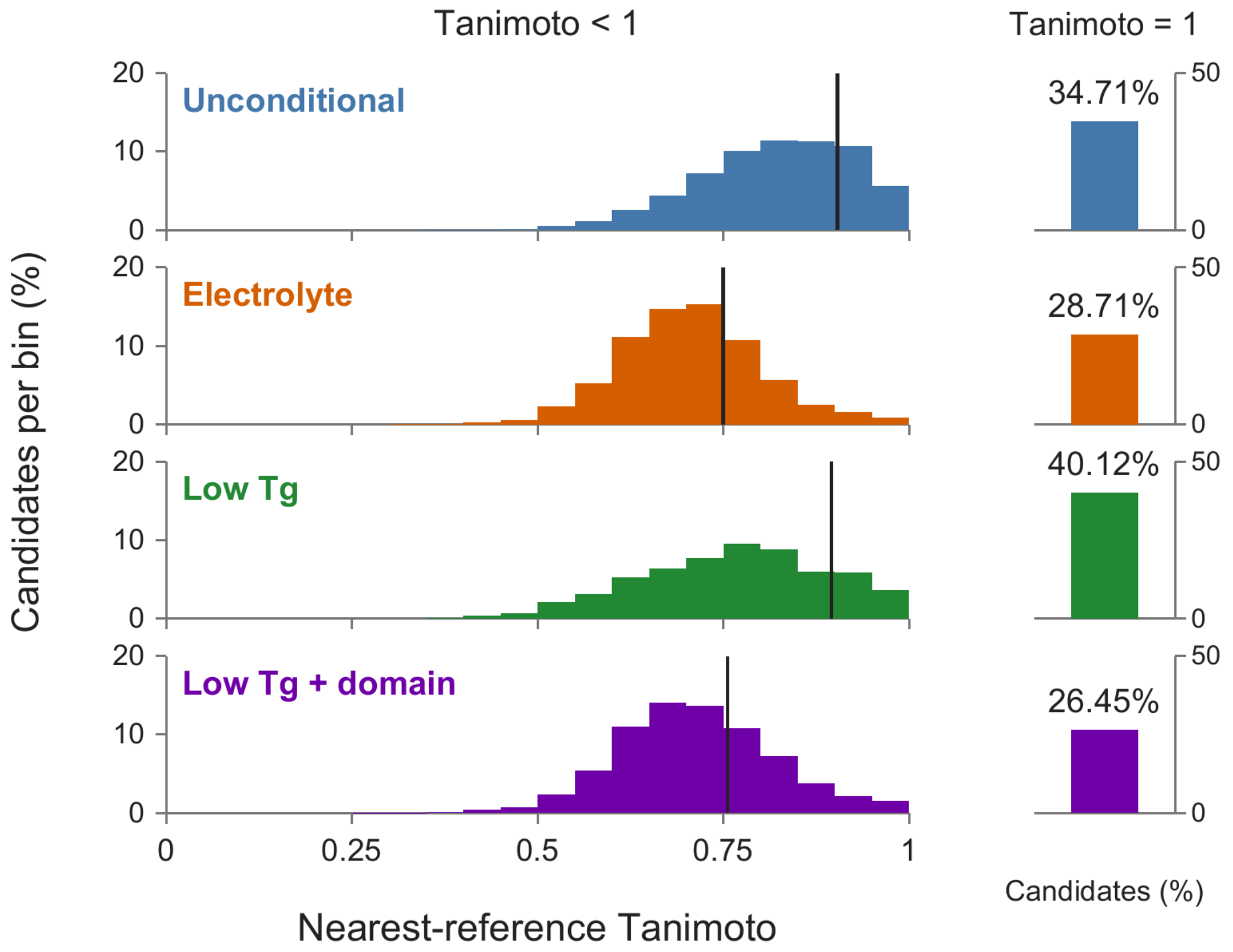

**Figure S7.** Nearest-neighbour similarity to encountered training structures. Each histogram shows maximum radius-two Morgan Tanimoto similarities below one in bins of width 0.05. Adjacent bars show exact-one fractions on their separately labelled percentage scale. Both use all distinct valid candidates in each condition as their denominator. Black lines mark overall medians, including exact-one values. Fingerprint equality does not establish canonical graph identity or memorization.

**S7 Structural control with electrolyte conditioning**

Structural prompts used the annotations in Table S3 and had the form BOS <task:generate> [<domain:electrolyte>] <must> <label> <target> or the corresponding <exclude> request. The electrolyte token was omitted for matched single-label controls. No label pairs or simultaneous inclusion and exclusion requests were used. Inclusion and exclusion results are reported in Tables S26 and S27, respectively.

Generation and validity rules followed Section S1.5. All four request variants for a label and seed were generated on a single GPU, with a rotating assignment of two seeds to the RTX 5070 Ti and one to the RTX 2000 Ada. CPU/GPU agreement was verified for all prompts on both devices before generation. To estimate how frequently each structural motif appears without an explicit structural request, we analyzed the first 2,048 matched attempts from the unconditional and electrolyte-only libraries.

Annotation failures remain unknown, not negative labels or confirmed instruction failures. Conditional compliance divides valid compliant draws by valid draws with known annotations. Distinct compliant yield counts each canonical compliant representation once across all three seeds and divides by all attempts; novel distinct compliant yield additionally requires verified absence from the complete eligible training catalog. Resolved periodic counts and unresolved identities are retained in the accompanying CSV.

**Table S26.** Single-label inclusion with and without electrolyte conditioning. Validity, compliant yield and distinct compliant yield are percentages of all attempts. Conditional compliance uses only valid outputs with known annotations; unknown is the number of valid draws with failed annotation. Distinct counts pool and deduplicate the three seeds.

| Label | Domain | Valid (%) | Unknown (n) | Conditional compliance (%) | Compliant yield (%) | Distinct compliant yield (%) |
|---|---|---|---|---|---|---|
| polyether | No | 99.53 | 5 | 91.80 | 91.29 | 91.28 |
| polyether | Yes | 99.59 | 0 | 5.20 | 5.18 | 4.49 |
| polycarbonate | No | 99.76 | 2 | 99.45 | 99.17 | 97.93 |
| polycarbonate | Yes | 99.67 | 0 | 96.49 | 96.18 | 67.12 |
| sulfone | No | 99.46 | 4 | 93.84 | 93.28 | 93.25 |
| sulfone | Yes | 99.46 | 2 | 0.61 | 0.60 | 0.41 |
| ester | No | 99.53 | 5 | 90.80 | 90.30 | 90.30 |
| ester | Yes | 99.48 | 4 | 7.27 | 7.23 | 6.14 |
| polysiloxane | No | 99.71 | 1 | 99.31 | 99.01 | 98.55 |
| polysiloxane | Yes | 99.69 | 3 | 0.05 | 0.05 | 0.05 |
| amide | No | 99.53 | 10 | 93.91 | 93.31 | 93.29 |
| amide | Yes | 99.61 | 1 | 92.58 | 92.20 | 69.03 |
| imide | No | 99.45 | 12 | 95.21 | 94.50 | 94.48 |
| imide | Yes | 99.61 | 1 | 0.29 | 0.29 | 0.29 |
| ether | No | 99.53 | 10 | 91.19 | 90.61 | 90.59 |
| ether | Yes | 99.58 | 7 | 17.30 | 17.20 | 13.70 |
| amine | No | 99.51 | 9 | 89.96 | 89.39 | 89.39 |
| amine | Yes | 99.58 | 2 | 51.90 | 51.66 | 39.26 |
| polyphosphazene | No | 99.84 | 1 | 99.98 | 99.80 | 96.89 |
| polyphosphazene | Yes | 99.69 | 1 | 0.00 | 0.00 | 0.00 |
| thioether | No | 99.66 | 2 | 96.65 | 96.29 | 96.22 |
| thioether | Yes | 99.22 | 0 | 73.93 | 73.36 | 27.05 |

**Table S27.** Single-label exclusion with and without electrolyte conditioning. Validity, compliant yield and distinct compliant yield are percentages of all attempts. Conditional compliance uses only valid outputs with known annotations; unknown is the number of valid draws with failed annotation. Distinct counts pool and deduplicate the three seeds.

| Label | Domain | Valid (%) | Unknown (n) | Conditional compliance (%) | Compliant yield (%) | Distinct compliant yield (%) |
|---|---|---|---|---|---|---|
| polyether | No | 99.64 | 12 | 96.74 | 96.21 | 96.21 |
| polyether | Yes | 99.51 | 0 | 94.91 | 94.45 | 77.05 |
| polycarbonate | No | 99.48 | 11 | 99.57 | 98.88 | 98.88 |
| polycarbonate | Yes | 99.51 | 0 | 38.50 | 38.31 | 32.31 |
| sulfone | No | 99.59 | 10 | 98.54 | 97.98 | 97.98 |
| sulfone | Yes | 99.50 | 1 | 99.53 | 99.01 | 81.22 |
| ester | No | 99.37 | 4 | 96.38 | 95.70 | 95.70 |
| ester | Yes | 99.54 | 3 | 92.85 | 92.38 | 74.43 |
| polysiloxane | No | 99.50 | 6 | 99.74 | 99.14 | 99.14 |
| polysiloxane | Yes | 99.56 | 0 | 99.98 | 99.54 | 80.76 |
| amide | No | 99.58 | 11 | 96.20 | 95.62 | 95.62 |
| amide | Yes | 99.19 | 4 | 87.93 | 87.16 | 56.69 |
| imide | No | 99.43 | 18 | 97.24 | 96.40 | 96.40 |
| imide | Yes | 99.66 | 4 | 99.85 | 99.45 | 80.35 |
| ether | No | 99.56 | 11 | 96.87 | 96.27 | 96.27 |
| ether | Yes | 99.35 | 2 | 83.55 | 82.98 | 67.43 |
| amine | No | 99.37 | 17 | 96.58 | 95.70 | 95.70 |
| amine | Yes | 99.67 | 3 | 54.27 | 54.07 | 45.20 |
| polyphosphazene | No | 99.61 | 3 | 99.92 | 99.48 | 99.48 |
| polyphosphazene | Yes | 99.59 | 4 | 100.00 | 99.53 | 80.62 |
| thioether | No | 99.61 | 11 | 99.08 | 98.52 | 98.52 |
| thioether | Yes | 99.54 | 2 | 96.93 | 96.45 | 78.69 |

**Table S28.** Novel distinct compliant yield and paired effects of the domain token. Novel yields use all attempts per condition and confirmed absence from the complete eligible training catalog. Differences are domain minus no-domain, in percentage points. Paired differences are calculated separately for each of the three seeds and summarized as mean plus or minus sample SD.

| Label | Request | Novel yield no domain (%) | Novel yield domain (%) | Paired distinct yield difference (pp) | Paired novel yield difference (pp) |
|---|---|---|---|---|---|
| polyether | Include | 36.69 | 2.93 | -86.34 ± 1.00 | -33.66 ± 0.33 |
| polyether | Exclude | 42.58 | 34.55 | -8.74 ± 0.74 | -6.61 ± 1.59 |
| polycarbonate | Include | 31.59 | 28.55 | -16.06 ± 0.34 | -0.81 ± 1.22 |
| polycarbonate | Exclude | 43.44 | 16.65 | -63.04 ± 0.75 | -26.12 ± 1.65 |
| sulfone | Include | 43.46 | 0.16 | -92.77 ± 0.42 | -43.29 ± 0.85 |
| sulfone | Exclude | 42.25 | 38.00 | -6.35 ± 0.59 | -2.41 ± 1.40 |
| ester | Include | 42.09 | 3.39 | -83.54 ± 0.38 | -38.61 ± 0.74 |
| ester | Exclude | 40.46 | 33.53 | -10.92 ± 1.24 | -5.42 ± 1.22 |
| polysiloxane | Include | 23.10 | 0.05 | -98.81 ± 0.16 | -23.08 ± 0.72 |
| polysiloxane | Exclude | 43.96 | 36.31 | -7.58 ± 0.69 | -5.91 ± 0.51 |
| amide | Include | 40.49 | 34.34 | -11.10 ± 1.86 | -3.40 ± 1.90 |
| amide | Exclude | 45.23 | 26.77 | -23.81 ± 0.29 | -15.79 ± 1.44 |
| imide | Include | 38.96 | 0.29 | -94.21 ± 0.23 | -38.67 ± 0.44 |
| imide | Exclude | 44.92 | 36.04 | -4.65 ± 1.12 | -7.06 ± 2.28 |
| ether | Include | 37.94 | 7.41 | -74.87 ± 0.86 | -30.18 ± 1.04 |
| ether | Exclude | 43.18 | 29.93 | -19.51 ± 1.18 | -12.00 ± 2.35 |
| amine | Include | 45.80 | 17.15 | -42.89 ± 0.64 | -27.47 ± 1.87 |
| amine | Exclude | 40.93 | 20.72 | -45.51 ± 0.86 | -19.50 ± 0.63 |
| polyphosphazene | Include | 13.33 | 0.00 | -98.67 ± 0.20 | -13.35 ± 0.61 |
| polyphosphazene | Exclude | 43.38 | 36.04 | -7.78 ± 0.61 | -5.50 ± 1.39 |
| thioether | Include | 46.84 | 20.87 | -56.82 ± 0.12 | -20.23 ± 0.12 |
| thioether | Exclude | 43.12 | 35.40 | -9.28 ± 0.76 | -6.01 ± 1.97 |

**Table S29.** Electrolyte training support and natural annotation prevalence without a structural request. Positive variants and groups count the 6,054 training variants and 5,808 groups; 12 variants in 12 groups have unknown annotations. A group can contain differently annotated variants, so positive- and negative-containing group counts can overlap. Baseline prevalence uses matched prefixes from the unconditional and electrolyte-only libraries. Only valid draws with known annotations enter the prevalence denominator, including repeated draws.

| Label | Positive variants | Positive groups | Negative groups | Natural prevalence no domain (%) | Natural prevalence domain (%) |
|---|---|---|---|---|---|
| polyether | 228 | 223 | 5573 | 28.35 | 3.92 |
| polycarbonate | 3666 | 3485 | 2311 | 0.33 | 62.09 |
| sulfone | 36 | 36 | 5760 | 9.75 | 0.87 |
| ester | 336 | 329 | 5467 | 9.52 | 6.14 |
| polysiloxane | 1 | 1 | 5795 | 0.58 | 0.02 |
| amide | 3575 | 3462 | 2334 | 62.02 | 55.64 |
| imide | 4 | 4 | 5792 | 54.88 | 0.26 |
| ether | 775 | 758 | 5038 | 34.65 | 14.82 |
| amine | 2399 | 2284 | 3512 | 22.58 | 44.69 |
| polyphosphazene | 1 | 1 | 5795 | 0.10 | 0.00 |
| thioether | 236 | 229 | 5567 | 3.54 | 3.46 |

We interpreted exclusion results in light of how often each motif appeared without a structural request (Table S29). Excluding a motif that was already rare provides less evidence that the model followed the instruction. High inclusion compliance can coincide with repeated generation of the same structures; distinct and novel distinct yields account for these duplicates. Although compliance tended to vary with how well a motif was represented in training, these comparisons do not establish that training frequency caused the differences. Different canonical repeat units may also represent the same periodic polymer.

**Table S30.** Sampling variation for every structural/domain request, expressed as mean plus or minus sample SD across three seeds. All three yields use attempts as the denominator. Each seed is deduplicated separately, so these means differ from pooled distinct yields in Tables S26-S28. These are sampling replicates of one trained model, not three independent training runs.

| Label | Request | Domain | Compliant yield (%) | Distinct compliant yield (%) | Novel distinct compliant yield (%) |
|---|---|---|---|---|---|
| polyether | Include | No | 91.29 ± 0.55 | 91.28 ± 0.52 | 36.69 ± 0.58 |
| polyether | Include | Yes | 5.18 ± 0.61 | 4.93 ± 0.56 | 3.03 ± 0.26 |
| polyether | Exclude | No | 96.21 ± 0.28 | 96.21 ± 0.28 | 42.58 ± 0.75 |
| polyether | Exclude | Yes | 94.45 ± 0.56 | 87.47 ± 0.76 | 35.97 ± 0.89 |
| polycarbonate | Include | No | 99.17 ± 0.17 | 98.67 ± 0.37 | 31.61 ± 0.07 |
| polycarbonate | Include | Yes | 96.18 ± 0.20 | 82.60 ± 0.68 | 30.79 ± 1.25 |
| polycarbonate | Exclude | No | 98.88 ± 0.21 | 98.88 ± 0.21 | 43.44 ± 1.16 |
| polycarbonate | Exclude | Yes | 38.31 ± 0.54 | 35.84 ± 0.85 | 17.32 ± 0.49 |
| sulfone | Include | No | 93.28 ± 0.35 | 93.28 ± 0.35 | 43.46 ± 0.79 |
| sulfone | Include | Yes | 0.60 ± 0.10 | 0.50 ± 0.16 | 0.16 ± 0.10 |
| sulfone | Exclude | No | 97.98 ± 0.24 | 97.98 ± 0.24 | 42.25 ± 0.88 |
| sulfone | Exclude | Yes | 99.01 ± 0.18 | 91.63 ± 0.42 | 39.84 ± 0.76 |
| ester | Include | No | 90.30 ± 0.28 | 90.30 ± 0.28 | 42.09 ± 0.77 |
| ester | Include | Yes | 7.23 ± 0.22 | 6.75 ± 0.29 | 3.48 ± 0.03 |
| ester | Exclude | No | 95.70 ± 0.38 | 95.70 ± 0.38 | 40.46 ± 0.40 |
| ester | Exclude | Yes | 92.38 ± 1.13 | 84.78 ± 1.00 | 35.04 ± 1.60 |
| polysiloxane | Include | No | 99.01 ± 0.17 | 98.86 ± 0.11 | 23.13 ± 0.73 |
| polysiloxane | Include | Yes | 0.05 ± 0.05 | 0.05 ± 0.05 | 0.05 ± 0.05 |
| polysiloxane | Exclude | No | 99.14 ± 0.14 | 99.14 ± 0.14 | 43.96 ± 0.33 |
| polysiloxane | Exclude | Yes | 99.54 ± 0.10 | 91.55 ± 0.55 | 38.05 ± 0.33 |
| amide | Include | No | 93.31 ± 0.60 | 93.29 ± 0.62 | 40.49 ± 0.90 |
| amide | Include | Yes | 92.20 ± 0.98 | 82.19 ± 1.24 | 37.09 ± 1.41 |
| amide | Exclude | No | 95.62 ± 0.07 | 95.62 ± 0.07 | 45.23 ± 1.13 |
| amide | Exclude | Yes | 87.16 ± 0.99 | 71.81 ± 0.33 | 29.44 ± 0.34 |
| imide | Include | No | 94.50 ± 0.41 | 94.50 ± 0.41 | 38.96 ± 0.56 |
| imide | Include | Yes | 0.29 ± 0.20 | 0.29 ± 0.20 | 0.29 ± 0.20 |

| Label | Request | Domain | Compliant yield (%) | Distinct compliant yield (%) | Novel distinct compliant yield (%) |
|---|---|---|---|---|---|
| imide | Exclude | No | 96.40 ± 0.66 | 96.40 ± 0.66 | 44.92 ± 1.90 |
| imide | Exclude | Yes | 99.45 ± 0.10 | 91.75 ± 0.69 | 37.86 ± 0.39 |
| ether | Include | No | 90.61 ± 0.33 | 90.61 ± 0.33 | 37.94 ± 1.14 |
| ether | Include | Yes | 17.20 ± 0.88 | 15.74 ± 0.56 | 7.76 ± 0.21 |
| ether | Exclude | No | 96.27 ± 0.20 | 96.27 ± 0.20 | 43.18 ± 0.59 |
| ether | Exclude | Yes | 82.98 ± 0.62 | 76.76 ± 1.14 | 31.18 ± 1.76 |
| amine | Include | No | 89.39 ± 0.45 | 89.39 ± 0.45 | 45.80 ± 0.61 |
| amine | Include | Yes | 51.66 ± 0.86 | 46.50 ± 0.94 | 18.33 ± 1.26 |
| amine | Exclude | No | 95.70 ± 0.35 | 95.70 ± 0.35 | 40.93 ± 0.42 |
| amine | Exclude | Yes | 54.07 ± 0.57 | 50.20 ± 0.62 | 21.44 ± 0.34 |
| polyphosphazene | Include | No | 99.80 ± 0.05 | 98.67 ± 0.20 | 13.35 ± 0.61 |
| polyphosphazene | Include | Yes | 0.00 ± 0.00 | 0.00 ± 0.00 | 0.00 ± 0.00 |
| polyphosphazene | Exclude | No | 99.48 ± 0.31 | 99.48 ± 0.31 | 43.38 ± 0.84 |
| polyphosphazene | Exclude | Yes | 99.53 ± 0.20 | 91.70 ± 0.75 | 37.87 ± 2.20 |
| thioether | Include | No | 96.29 ± 0.42 | 96.24 ± 0.43 | 46.84 ± 0.65 |
| thioether | Include | Yes | 73.36 ± 2.13 | 39.42 ± 0.42 | 26.61 ± 0.73 |
| thioether | Exclude | No | 98.52 ± 0.31 | 98.52 ± 0.31 | 43.12 ± 1.10 |
| thioether | Exclude | Yes | 96.45 ± 0.29 | 89.24 ± 0.94 | 37.11 ± 1.01 |

## References

[1] Kuenneth, Christopher; Ramprasad, Rampi. polyBERT: a chemical language model to enable fully machine-driven ultrafast polymer informatics. Nature Communications 2023, 14, 4099. https://doi.org/10.1038/s41467-023-39868-6

[2] Kuenneth, Christopher; Ramprasad, Rampi. polyOne Data Set - 100 million hypothetical polymers including 29 properties. Zenodo 2022, version 1. https://doi.org/10.5281/zenodo.7124188

[3] Bradford, Gabriel; Lopez, Jeffrey; Ruza, Jurgis; Stolberg, Michael A.; Osterude, Richard; Johnson, Jeremiah A.; Gomez-Bombarelli, Rafael; Shao-Horn, Yang. Chemistry-Informed Machine Learning for Polymer Electrolyte Discovery. ACS Central Science 2023, 9 (2), 206-216. https://doi.org/10.1021/acscentsci.2c01123

[4] Schauser, Nicole S.; Kliegle, Gabrielle A.; Cooke, Piper; Segalman, Rachel A.; Seshadri, Ram. Database Creation, Visualization, and Statistical Learning for Polymer Li+-Electrolyte Design. Chemistry of Materials 2021, 33 (13), 4863-4876. https://doi.org/10.1021/acs.chemmater.0c04767

[5] Hatakeyama-Sato, Kan; Tezuka, Toshiki; Umeki, Momoka; Oyaizu, Kenichi. AI-Assisted Exploration of Superionic Glass-Type Li+ Conductors with Aromatic Structures. Journal of the American Chemical Society 2020, 142 (7), 3301-3305. https://doi.org/10.1021/jacs.9b11442

[6] Yang, Zhenze; Ye, Weike; Lei, Xiangyun; Schweigert, Daniel; Kwon, Ha-Kyung; Khajeh, Arash. De novo design of polymer electrolytes using GPT-based and diffusion-based generative models. npj Computational Materials 2024, 10, 296. https://doi.org/10.1038/s41524-024-01470-9

[7] Xie, Tian; France-Lanord, Arthur; Wang, Yanming; Lopez, Jeffrey; Stolberg, Michael A.; Hill, Megan; Leverick, Graham Michael; Gomez-Bombarelli, Rafael; Johnson, Jeremiah A.; Shao-Horn, Yang; Grossman, Jeffrey C. Accelerating amorphous polymer electrolyte screening by learning to reduce errors in molecular dynamics simulated properties. Nature Communications 2022, 13, 3415. https://doi.org/10.1038/s41467-022-30994-1

[8] Xie, Tian; Kwon, Ha-Kyung; Schweigert, Daniel; Gong, Sheng; France-Lanord, Arthur; Khajeh, Arash; Crabb, Emily; Puzon, Michael; Fajardo, Chris; Powelson, Will; Shao-Horn, Yang; Grossman, Jeffrey C. A cloud platform for sharing and automated analysis of raw data from high throughput polymer MD simulations. APL Machine Learning 2023, 1 (4), 046108. https://doi.org/10.1063/5.0160937

[9] Ruza, Jurgis; Stolberg, Michael. Experimental data for "Autonomous Discovery of Polymer Electrolyte Formulations with Warm-start Batch Bayesian Optimization". Zenodo 2025, version v1. https://doi.org/10.5281/zenodo.16741235

[10] Stolberg, Michael A.; Lopez, Jeffrey; Cawthern, Sawyer D.; Herzog-Arbeitman, Abraham; Kwon, Ha-Kyung; Schweigert, Daniel; Anapolosky, Abraham; Storey, Brian D.; Johnson, Jeremiah A.; Shao-Horn, Yang. A data-driven platform for automated characterization of polymer electrolytes. Matter 2025, 8 (6), 102129. https://doi.org/10.1016/j.matt.2025.102129

[11] Xu, Changwen; Wang, Yuyang; Barati Farimani, Amir. TransPolymer: a Transformer-based language model for polymer property predictions. npj Computational Materials 2023, 9, 64. https://doi.org/10.1038/s41524-023-01016-5

[12] Kuenneth, Christopher; Rajan, Arunkumar Chitteth; Tran, Huan; Chen, Lihua; Kim, Chiho; Ramprasad, Rampi. Polymer informatics with multi-task learning. Patterns 2021, 2 (4), 100238. https://doi.org/10.1016/j.patter.2021.100238

[13] linyeping. Extra dataset with SMILES,$T_g$,PID,Polimers Class. Kaggle dataset. https://www.kaggle.com/datasets/linyeping/extra-dataset-with-smilestgpidpolimers-class

[14] Tang, Hao; Yue, Tianle; Li, Ying. Assessing Uncertainty in Machine Learning for Polymer Property Prediction: A Benchmark Study. Journal of Chemical Information and Modeling 2025, 65 (13), 6585-6598. https://doi.org/10.1021/acs.jcim.5c00550

[15] Jupiterj. polyLearn. GitHub repository, revision e4c20cb59e45. https://github.com/Jupiterj/polyLearn

[16] WangGroupFDU. Openpoly-benchmark. GitHub repository, revision ccaa66900747. https://github.com/WangGroupFDU/Openpoly-benchmark

[17] BoranLilike. PolyFKAN: Integrating Fourier-KAN Layers into Graph Neural Networks for Polymer Property Prediction. GitHub repository, revision 83dcafb32126. https://github.com/BoranLilike/PolyFKAN

[18] Otsuka, Shingo; et al. PoLyInfo: Polymer Database for Polymeric Materials Design. In 2011 International Conference on Emerging Intelligent Data and Web Technologies; 2011; pp 22-29. https://doi.org/10.1109/EIDWT.2011.13